\documentclass[usenatbib]{mnras}

\usepackage[T1]{fontenc}
\usepackage{ae,aecompl}

\usepackage{graphicx}	
\usepackage{amsmath}	
\usepackage{amssymb}	
\usepackage{booktabs}
\usepackage{eso-pic}

\usepackage{ulem}       
\usepackage{url}
\usepackage{times}

\def \cf2{{\it Cosmicflows-2\,}}

\def \der{{\rm d}}
\def \msun{{{\rm M}_{\odot}}}

\def \araa{Annual Review of Astron and Astrophys}
\def\mnras{Monthly Notices of the Royal Astronomical Society}
\def\apj{The Astrophysical Journal}
\def\apjl{Astrophysical Journal Letters}
\def\aap{Astronomy and Astrophysics}
\def\physrep{Physics Reports}
\def\jcap{Journal of Cosmology and Astroparticle Physics}
\def\prd{Physical Review D}

\def\nat{Nature}

\def\apjs{The Astrophysical Journal Supplement Series}
\def\aj{The Astronomical Journal}

\makeatletter
\let\ftype@table\ftype@figure
\makeatother

\title[Cluster Pressure Profile]{Probing the cluster pressure profile with thermal Sunyaev-Zeldovich effect and weak lensing cross-correlation}
\author[Ma et al.]{Yin-Zhe Ma$^{1,2,3,\dagger}$, Yan Gong$^{4,3,\star}$, Tilman Tr\"{o}ster$^{5}$, Ludovic Van Waerbeke$^{6}$ \\
$^{1}$ Purple Mountain Observatory, CAS, No.8 Yuanhua Road, Qixia District, Nanjing 210034, China \\
$^{2}$ School of Chemistry and Physics, University of KwaZulu-Natal, Westville Campus, Private Bag X54001, Durban, 4000, South Africa \\
$^{3}$ NAOC-UKZN Computational Astrophysics Centre (NUCAC), University of KwaZulu-Natal, Durban, 4000, South Africa \\
$^{4}$ Key Laboratory of Space Astronomy and Technology, National Astronomical Observatories, Chinese Academy of Sciences, Beijing 100012, China \\
$^{5}$ Institute for Astronomy, University of Edinburgh, Royal Observatory, Blackford Hill, Edinburgh, EH9 3HJ, UK \\
$^{6}$ Department of Physics and Astronomy, University of British Columbia, 6224 Agricultural Road, Vancouver, BC, V6T 1Z1, Canada \\
{\rm Emails}: $^{\dagger}$ma@ukzn.ac.za \\
$^{\star}$gongyan@bao.ac.cn \\
}

\pubyear{2018}

\begin{document}
\label{firstpage}
\pagerange{\pageref{firstpage}--\pageref{lastpage}}
\maketitle

\begin{abstract}
We confront the universal pressure profile (UPP) proposed by~\citet{Arnaud10} with the recent measurement of the cross-correlation function of the thermal Sunyaev-Zeldovich (tSZ) effect from {\it Planck} and weak gravitational lensing measurement from the Red Cluster Sequence lensing survey (RCSLenS). By using the halo model, we calculate the prediction of $\xi^{y-\kappa}$ (lensing convergence and Compton-$y$ parameter) and  $\xi^{y-\gamma_{\rm t}}$ (lensing shear and Compton-$y$ parameter) and fit the UPP parameters by using the observational data. We find consistent UPP parameters when fixing the cosmology to either {\it WMAP} 9-year or {\it Planck} 2018 best-fitting values. The best constrained parameter is the pressure profile concentration $c_{500}=r_{500}/r_{\rm s}$, for which we find $c_{500} = 2.68^{+1.46}_{-0.96}$ ({\it WMAP}-9) and $c_{500} = 1.91^{+1.07}_{-0.65}$ ({\it Planck}-2018) for the $\xi^{y-\gamma_t}$ estimator. The shape index for the intermediate radius region $\alpha$ parameter is constrained to $\alpha=1.75^{+1.29}_{-0.77}$ and $\alpha = 1.65^{+0.74}_{-0.5}$ for {\it WMAP}-9 and {\it Planck}-2018 cosmologies, respectively. Propagating the uncertainties of the UPP parameters to pressure profiles results in a factor of $3$ uncertainty in the shape and magnitude. Further investigation shows that most of the signal of the cross-correlation comes from the low-redshift, inner halo profile ($r \leqslant r_{\rm vir}/2$) with halo mass in the range of $10^{14}$--$10^{15}\msun$, suggesting that this is the major regime that constitutes the cross-correlation signal between weak lensing and tSZ.

\end{abstract}
%

%



\section{Introduction}
\label{sec:intro}

Galaxy clusters are essential objects in understanding the galaxy and structure formation. The clusters are normally filled with hot and warm ionised plasma that can be measured via X-ray imaging and the thermal Sunyaev-Zeldovich effect~\citep{Sunyaev72,Sunyaev80}. The intensity of the observed X-ray image depends on the square of the electron density profile$~n^{2}_{\rm e}$, which is more sensitive to the central hot baryons. The thermal Sunyaev-Zeldovich (tSZ) effect, in contrast, depends on the integration of the electron density and temperature. Therefore, for low mass haloes and filaments where the temperature is low, the systems can still have large values of $n_{\rm e}$ where most baryons reside. Therefore, the tSZ effect can trace down the warm baryons that are both in the centre of the massive halos and diffused outside the centre halos~\citep{Waerbeke14,Ma15} and associated with filamentary structures~\citep{Tanimura19,Graaff19}. There has been a growing interest in recent years to predict and measure the tSZ effect in radio and microwave observations~\citep{Birkinshaw78,Birkinshaw99,Carlstrom02,Ma15}.

The tSZ effect is a secondary anisotropy in the cosmic microwave background radiation (CMB), which is directly related to the pressure profile of galaxy clusters. The effect is caused by inverse Compton scattering of cosmic microwave background (CMB) photons by the hot plasma in clusters of galaxies. The temperature anisotropy caused by tSZ effect is
\begin{eqnarray}
\frac{\Delta T}{T_{\rm CMB}}
= \left[\eta \frac{e^{\eta}+1}{e^{\eta}-1} - 4\right] y \equiv
g_{\nu}y, \label{eq:deltaT1}
\end{eqnarray} 
where 
\begin{eqnarray}
g_{\nu} \equiv \eta\frac{{\rm e}^{\eta}+1}{{\rm e}^{\eta}-1} -4, \label{eq:gnu}
\end{eqnarray}
and $\eta=h\nu/k_{\rm B}T_{\rm CMB}=1.76\,(\nu/100\,{\rm GHz})$.
The dimensionless parameter $y$ is called Compton-$y$ parameter, defined as
\begin{eqnarray}
y = \int n_{\rm{e}}({\bf
r})\sigma_{\rm{T}} \,\frac{k_{\rm{B}} T_{\rm{e}}({\bf
r})}{m_{\rm{e}} c^2} \,\der l, \label{eq:comptony}
\end{eqnarray}   
where
where the integral is taken along the line-of-sight of the pressure profile. $m_{\rm{e}}$ is the electron rest mass, $k_{\rm{B}}$ is
the Boltzmann constant, and $\sigma_{\rm{T}}$ is the Thomson cross section. We treat the electrons as ideal gases. In this sense, the pressure profile can be defined as 
$P_{\rm e}(r)=n_{\rm e}k_{\rm B}T_{\rm e}$. By using maps from multi-frequency channels of the ESA's {\it Planck} survey, one can isolate the frequency factor (Eq.~(\ref{eq:gnu})) from the temperature fluctuation (Eq.~(\ref{eq:deltaT1})) and obtain a direct Compton-$y$ map~\citep{Waerbeke14,Planck-alksky14}. Thus by modelling of the pressure profile, one can compare it with the measured  Compton-$y$ map and constrain the profile. 

In practice, the measured all-sky Compton-$y$ map is quite noisy with thermal noise and residual foregrounds. The systematics remained in the Compton $y-$map is uncorrelated with the systematics in the RCSLenS map. Therefore, the cross-correlation study can efficiently extract the underlying baryons signal presented in both two surveys, which is hard to measure in separate studies. There have been several cross-correlation studies by using different large-scale structure tracers to extract the baryonic bias. These tracers resemble the underlying dark matter distribution, such as the CMB lensing~\citep{Hill13}, galaxy groups of different halo masses selected from Sloan Digital Sky Survey (SDSS) low-$z$ catalogue~\citep{Vikram17,Lim18}, and weak gravitational lensing measurement from Canada France Hawaii Telescope Lensing Survey (CFHTLenS)~\citep{Waerbeke14,Ma15,Hojjati15} and Red Cluster Sequence Lensing Survey (RCSLenS)~\citep{Hojjati17}. Each of the cross-correlation methods has its advantage and disadvantage. For instance, the cross-correlation with CMB lensing map can reach $6.2\sigma$ confidence level (C.L.) detection, but the CMB lensing kernel peaks at $z \simeq 2$, and receives significant contributions over a wide redshift range ($0.1<z<10$)~\citep{Hill13}. Therefore, such cross-correlation will not be very sensitive to the low-redshift ``missing baryon'' component\footnote{As an alternative approach, the dispersion measurement of the Fast Radio Bursts (FRB) is the line-of-sight integral of all ionized electrons from observer to the source, which essentially captures all baryon's signal~\citep{Macquart20}. The cross-correlation study between FRB and tSZ map also help revealing baryon distribution~\citep{Munoz18}.}. The cross-correlation between tSZ effect and galaxy groups also reaches a high-level significance~\citep{Lim18}, but it is quite sensitive to bias in between galaxy and dark matter distribution.

In this paper, we will focus on probing the galaxy cluster pressure profile from the cross-correlation of tSZ effect with optical lensing shear and convergence maps. The first detection of this cross-correlation was made by~\citet{Waerbeke14} at $\sim 6\sigma$ C.L., by using {\it Planck} nominal mission maps and $154\,{\rm deg}^{2}$ CFHT Lensing map. The follow-up studies showed that this correlation function receives a non-negligible contribution from baryons resided in low-mass halos, and baryons outside the virial radius of the halos which is hard to be detected by X-ray imaging surveys and galaxy groups correlation~\citep{Ma15,Hojjati15}. By conducting a similar route of study, \citet{Hojjati17} used the $560\,{\rm deg}^{2}$ RCSLenS map~\footnote{This is the effective area of the survey. The total area is $\sim 780\,{\rm deg}^{2}$~\citep{Hilderbrandt16}.} to cross-correlate with {\it Planck} full-mission $y-$map, and reported $7.1\sigma$ and $8.1\sigma$ detections of the cross-correlation using configuration-space $y-\kappa$ and $y-\gamma_{\rm t}$ estimators respectively. In this paper, we compute the halo model with universal pressure profile (UPP)~\citep{Arnaud10} for both {\it WMAP}-9 and {\it Planck} 2018 cosmological parameters, and confront the prediction of the $y-\kappa$ and $y-\gamma_{\rm t}$ correlation with the data measured in~\citet{Hojjati17}. Our goal is twofold. One is to put the UPP profile into the test with the current cross-correlation data and examine whether the halo concentration from lensing-SZ correlation is consistent with the previous finding by using cluster studies~(\citet{Arnaud10} and~\citet{Planck13-pressure}). The other is to investigate whether the 2-pt correlation function from the halo model can adequately describe the detected $\kappa$-$y$ cross-correlation from the data. 

This paper is organised as follows. In Sec.~\ref{sec:data}, we describe the RCSLenS map, {\it Planck} all-sky Compton-$y$ parameter map used in the correlation function, and the measured cross-correlation function $\xi^{y-\gamma_{\rm t}}$ and $\xi^{y-\kappa}$. In Sec.~\ref{sec:theory}, we first describe the UPP model, and then use the halo model to compute the theoretical $\xi^{y-\gamma_{\rm t}}$ and $\xi^{y-\kappa}$ correlation function. Then we present our likelihood analysis method. In Sec.~\ref{sec:results}, we present the results of our fitting and discuss the implication of the results. We present the conclusion remark and outlook in the last section.

\begin{table*}
\begin{centering}
\begin{tabular}{@{}llllllll}
\hline
Parameters & Reference & $\Omega_{\rm b}$ & $\Omega_{\rm m}$ & $\Omega_{\Lambda}$ & $h$ & $\sigma_{8}$ & $n_{\rm s}$ \\
 \hline
{\it Planck}-2018 & \citet{Planck-parameters18} & $0.0493$ & $0.315$ & $0.685$ & $0.674$ & $0.811$ & $0.965$ \\ \hline
{\it WMAP}-9 & \citet{Hinshaw13} & $0.0463$ & $0.279$ & $0.721$ & $0.70$ & $0.821$ & $0.972$ \\ \hline
\end{tabular}%
\caption{The cosmological parameters used in the numerical analysis in Sec.~\ref{sec:results}.}
\label{tab:cosmology}
\end{centering}
\end{table*}

Throughout the paper, we adopt a spatially-flat $\Lambda$CDM cosmology model with cosmological parameters defined as Hubble constant is $H_{0}=100h\,{\rm km}\,{\rm s}^{-1}\,{\rm Mpc}^{-1}$, fractional baryon (matter) density $\Omega_{\rm b}$ ($\Omega_{\rm m}$), {\it rms} fluctuation of matter power spectrum at $8\,h^{-1}$Mpc scale, and spectral index of the primordial power spectrum $n_{\rm s}$. In Sec.~\ref{sec:results}, the UPP model profile will be fitted for assumed {\it WMAP}-9 and {\it Planck}-2018 cosmological parameters, with values listed in Table~\ref{tab:cosmology}. 

\section{The Cross-correlation data}
\label{sec:data}

\begin{figure}
\centerline{\includegraphics[width=3.2in]{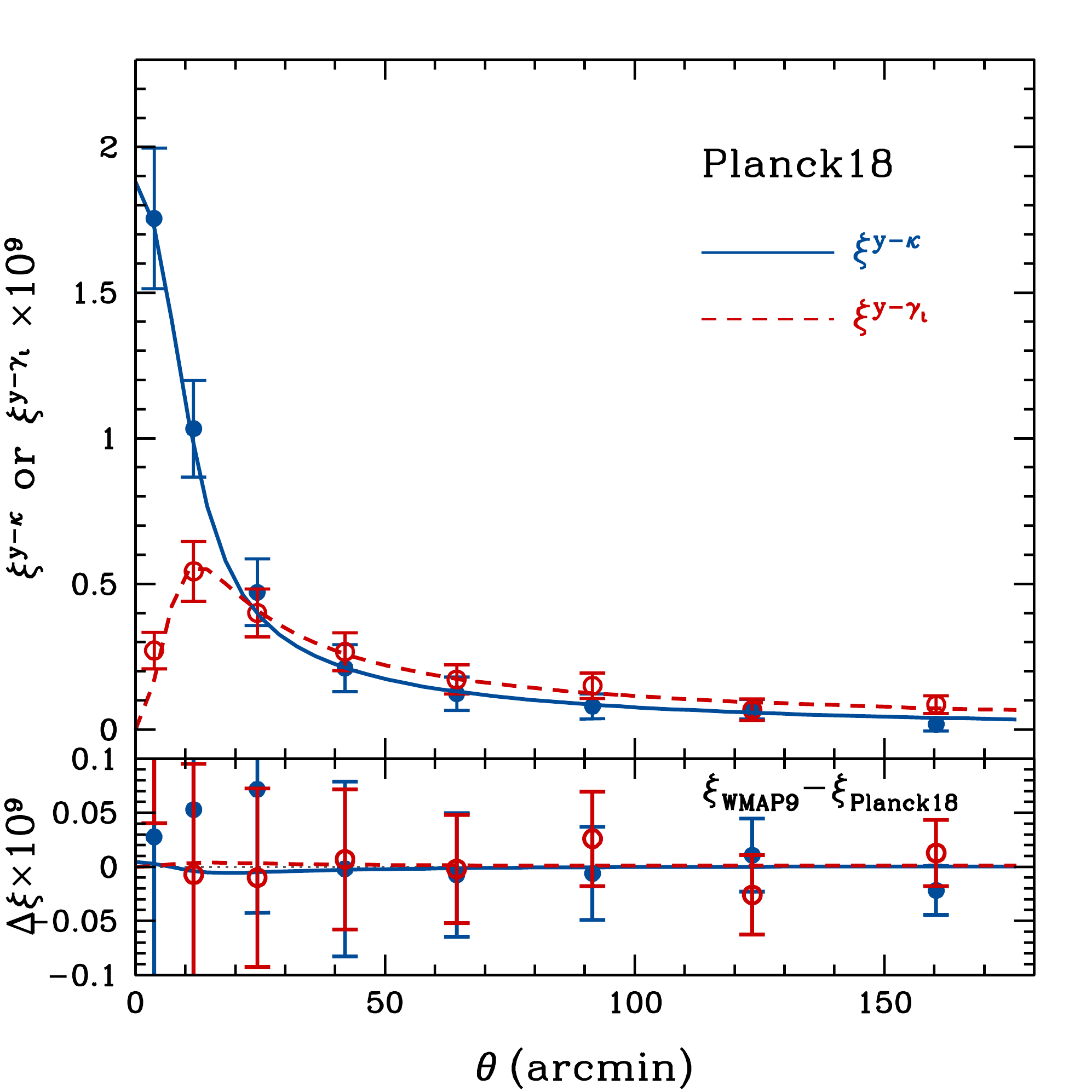}}
\caption{Comparison between the data and theory of $\xi^{y-\kappa}(\theta)$ and $\xi^{y-\gamma_{\rm t}}(\theta)$ with assumptions of {\it Planck}-2018 (upper panel) and {\it WMAP}-9 (lower panel showing the difference) cosmological parameters. The blue and red data are the $y$--$\kappa$ and $y$--$\gamma_{\rm t}$ cross-correlations respectively. The lower panel shows the difference between the best-fitting correlation functions of {\it Planck} and {\it WMAP} ($\Delta \xi \equiv \left(\xi_{\rm WMAP9}- \xi_{\rm Planck18}\right)\times 10^{9}$), which suggests that {\it WMAP} $\xi^{y-\kappa}(\theta)$ ($\xi^{y-\gamma_{\rm t}}(\theta)$) has slightly lower (higher) amplitude at small angular separation, but this decrement is negligible comparing to the error-bar of the data.} \label{fig:Planck-WMAP}
\end{figure}

The tSZ and lensing cross-correlation data we use here is the correlation between {\it Planck} map and the Red Cluster Sequence Lensing Survey (RCSLenS), shown in~\citet{Hojjati17} for detail. The {\it Planck} tSZ map is taken as the {\tt MILCA} map~\citep{Planck-XXII16} publicly available on {\it Planck} legacy Archive~\footnote{\url{https://pla.esac.esa.int/home}}. The {\tt MILCA} map has the Full-Width-Half-Maximum (FWHM) equals to $10$\,arcmin. There was no re-processing of the {\it Planck} tSZ data, other than cutting the patches that match the footprints of RCSLenS data.

The RCSLenS data was acquired from MegaCAM camera from $14$ separate fields and covered a total area of $785\,{\rm deg}^{2}$ on the sky~\citep{Hilde16,Gilbank11}. We obtained the data after using the reduction algorithm, photometric redshift estimation, and a shape measurement algorithm. For a complete treatment of the photometric data, please see~\citet{Heymans12} and~\citet{Hilde16} for details. After the magnitude cut of ${\rm mag}_{\rm r}>18$, the redshift distribution is shown in fig.~1 in~\citet{Hojjati17}, which peaks at $z=0.5$ but extends up to $z\simeq 2$.

For the cross-correlation, \citet{Hojjati17} used the shear data and reconstructed convergence maps for RCSLenS. The reconstruction of the convergence map is presented as in~\citet{Waerbeke14}. It is demonstrated that one can achieve the best SNR when the $\kappa$ map is smoothed at the same scale of {\it Planck} survey, i.e. $\theta_{\rm FWHM}=10\,$arcmin. The tSZ--tangential shear cross-correlation is obtained at the catalogue level where each pixel of the $y$-map is correlated with the average tangential shear from the corresponding shear data around the point~\citep{Hojjati17}. Therefore, the shear catalogue is not smoothed, so we should not include the smoothing kernel during the computation of theoretical correlation function. 

For the estimation of the covariance matrix in configuration space, the method in~\citet{Waerbeke14} is used. In brief, three hundred random catalogues from each of RCSLenS fields are created by randomly rotating the individual galaxies. This procedure will destroy the underlying lensing signal and create pure statistical noise. Each random catalogue is correlated with {\it Planck} tSZ map, and the total variance is calculated as the covariance matrix for $\xi^{y-\gamma_{\rm t}}$. These 300 random shear catalogues are also used to construct a set of convergence noise maps. Then the noise maps are correlated with the tSZ map to quantify the covariance of the  $\xi^{y-\kappa}$. Once we completed the above procedure, we found that there is a significant correlation out to $3^{\circ}$ of the angular separation on the sky. The signals correspond to $13\sigma$ and $17\sigma$ detections of the cross-correlations of the $\kappa-y$ and $\gamma_{\rm t}-y$ estimators respectively.

~\citet{Hojjati17} further included an estimate of the sampling variance into the total covariance matrix, because the observed fields are small and there is significant scatter in between these fields. We compared the variance in each angular bin to the reconstructed covariance matrix calculated before. We estimated the scaling factor by which one should inflate the computed covariance matrix to match the scatters between fields. We show the resultant cross-correlation data and the error-bar at each angular separation in Fig.~\ref{fig:Planck-WMAP}, with blue and red data with error-bars being $\xi^{y-\kappa}$ and $\xi^{y-\gamma_{\rm t}}$ respectively. The error-bar is the square-root of the diagonal elements in the covariance matrix, which are shown in fig.~4 in~\citet{Hojjati17}. One can see that the covariance matrix are quite symmetric along with the diagonal elements, indicating that the cross-angular correlation is quite small. By including the sample variance, the detections of $\xi^{y-\kappa}$ and $\xi^{y-\gamma_{\rm t}}$ are measured at $7.1\sigma$\,C.L. and $8.1\sigma$\,C.L. respectively.

\section{Theory}
\label{sec:theory}

\subsection{Compton-$y$ parameter profile}
The 2D tSZ signal in $\ell$-space is an integrated Fourier transformation of pressure profile
\begin{eqnarray}
y_{\ell}(M,z)=\frac{a(\chi)}{\chi^{2}(z)}\int^{\infty}_{0} \der r (4\pi
r^{2}) \frac{\sin(\ell r/\chi)}{\ell r/\chi} y_{\rm 3D}(r;M,z),
\label{eq:y3D1}
\end{eqnarray}
where $\chi$ is the comoving distance to redshift $z$, $r$ is
comoving radial distance of $y_{\rm 3D}$ profile. $y_{\rm 3D}$
is the profile with physical unit, which is~\citep{Hill13}
\begin{eqnarray}
y_{\rm 3D}(r;M,z)=\frac{\sigma_{\rm T}}{m_{\rm e}c^{2}}P_{\rm
e}\left(r;M,z\right).
\end{eqnarray}

In the Universal Pressure Profile $P_{\rm e}(r)$, $R_{500}$ is defined as the radius of the halo within which the density is $500$ times the critical density of the Universe, i.e. $M_{500}=(4\pi/3)500\rho_{\rm cr}(z)R^{3}_{500}$. With $x\equiv r/R_{500}$, the form of the universal profile of electron given
by~\citet{Arnaud10} is~\citep{Arnaud10,Planck13-pressure}
\begin{eqnarray}
P_{\rm e}(r)=P_{500}F(M_{500})\mathbb{P}(x=r/R_{500}),\label{eq:unipres}
\end{eqnarray}
where
\begin{eqnarray}
P_{500}&=& 1.65 \times 10^{-3}E^{8/3}(z) \nonumber \\
& \times &
 \left[\frac{M_{500}}{3\times 10^{14}\msun h^{-1}_{70}} 
\right]^{2/3} h^{2}_{70} [{\rm keV \textrm{ }cm^{-3}}], \label{eq:P500}
\end{eqnarray}
and
\begin{eqnarray}
F(M_{500})=\left[\frac{M_{500}}{3 \times 10^{14} \msun h^{-1}_{70}} \right]^{\alpha_{\rm p}},
\end{eqnarray}
in which $h_{70}=(h/0.7)$, $\alpha_{\rm{p}} =0.12$. The ``Universality'' of the pressure profile resembles in
the $\mathbb{P}(x)$ function, which is the generalized NFW model 
\begin{eqnarray}
\mathbb{P}(x) = \frac{P_0}{(c_{500} x)^{\gamma}\left[1+(c_{500}
x)^{\alpha}\right]^{(\beta-\gamma)/\alpha}}, \label{eq:px}
\end{eqnarray} 
where $P_{0}$ is the overall
magnitude of the pressure profile, $c_{500}$ is the pressure profile concentration parameter, and $\gamma, \alpha$,
and $\beta$ determine the slope of the profile. As one can see from the equation, for small radius ($x\rightarrow 0$), $\mathbb{P}(x) \rightarrow x^{-\gamma}$; and for large radius ($x\rightarrow \infty$), $\mathbb{P}(x) \rightarrow x^{-\beta}$. Therefore, $\gamma$ and $\beta$ more or less determine the inner and outer slopes of the profile (note that $c_{500}$ also affects), and the $\alpha$ affects the transition in between. Previous studies using different tracers of the galaxy clusters found different results of the fitting. For instance, ~\citet{Arnaud10} used $33$ local ($z<0.2$) massive ($10^{14}\msun<M_{500}<10^{15}\msun$) for the fitting,~\citet{Planck13-pressure} used $62$ massive clusters ($2 \times 10^{14}\msun<M_{500}<2 \times 10^{15}\msun$) to derive the parameters of UPP, and more recently, ~\citet{Gong19} used $101,407$ locally ($0.16<z<0.47$) most-massive luminous red galaxies to probe the UPP profile. We list the results of the above parameter fitting in Table~\ref{tab-para} for comparison.


\begin{table*}
\begin{centering}
\begin{tabular}{@{}lllllll}
\hline References & $P_{0}$ & $c_{500}$ & $\alpha$ & $\beta$ & $\gamma$ & $\chi^{2}_{\rm min}$ \\
\hline \citet{Arnaud10} & $8.403 h^{-3/2}_{70}$ & $1.177$ &
$1.0510$ & $5.4905$ & $0.31$ & N/A \\ \hline ~\citet{Planck13-pressure} &
$6.41$ & $1.81$ & $1.33$ & $4.13$ & $0.31$ & N/A \\ \hline 
\citet{Gong19} & $2.99^{+3.44}_{-1.57}$ & $1.16^{+0.79}_{-0.29}$ & $2.66^{+1.67}_{-0.97}$ & $5.48^{+2.39}_{-1.38}$ & $0.31$ & N/A \\ \hline
This work ({\it WMAP}-9 $\xi^{y-\kappa}$) & $17.93^{+2.07}_{-14.47} $ & $2.69^{+0.85}_{-0.97}$ & $2.15^{+6.16}_{-1.06} $ & $3.46^{+1.49}_{-0.62}$ & $0.31$ & $2.83$ \\ \hline
This work ({\it Planck}-2018 $\xi^{y-\kappa}$) & $9.68^{+10.02}_{-7.11}$ & $2.71^{+0.92}_{-0.93} $ & $5.97^{+1.81}_{-4.73}$ & $3.47^{+1.39}_{-0.60} $ & $0.31$ & $2.72$ \\ \hline 
This work ({\it WMAP}-9 $\xi^{y-\gamma_{\rm t}}$) & $16.91^{+3.05}_{-11.29}$ & $2.68^{+1.46}_{-0.96}$ & $1.75^{+1.29}_{-0.77}$ & $2.78^{+0.53}_{-0.37}$ &$0.31$ & $6.56$ \\ \hline
This work ({\it Planck}-2018 $\xi^{y-\gamma_{\rm t}}$) & $6.62^{+2.06}_{-1.65}$ & $1.91^{+1.07}_{-0.65}$ & $1.65^{+0.74}_{-0.50}$ & $4.88^{+1.18}_{-2.46}$ & $0.31$ & $6.54$ \\ \hline 
\end{tabular}%
\caption{The best-fitting UPP parameters previous studies and this work, with fixed value of $\gamma=0.31$. The minimal $\chi^{2}$ values are listed for this work ($N_{\rm dof}=8-4=4$). The data used in~\citet{Arnaud10} is $33$ local ($z<0.2$) clusters drawn from the REFLEX catalogue and observed with XMM-Newton with mass in range $10^{14}\msun < M_{500} < 10^{15} \msun$, and in~\citet{Planck13-pressure} is $62$ local ($z<0.5$ but mostly $z<0.3$) clusters with mass in range $10^{14}\msun < M_{500} < 1.5 \times 10^{15} \msun$. The best-fitting values quoted in~\citet{Arnaud10} and~\citet{Planck13-pressure} do not contain the confidence intervals possibly due to the small number of statistics, though~\citet{Planck13-pressure} shows the likelihoods and joint constraints of these parameters. The major difference between this work and the previous literatures is that this work uses the weak-lensing and tSZ cross-correlation function to constrain these parameters, instead of using cluster samples.}
\label{tab-para}
\end{centering}
\end{table*}

\subsection{Weak Lensing}

The lensing convergence is an integral of 3-D over-density along the line-of-sight~\citep{Waerbeke14}
\begin{eqnarray}
\kappa(\hat{\mathbf{n}})=\int^{\chi_{\infty}}_{0}\der \chi
W^{\kappa}(\chi) \delta_{\rm m, 3D}(\chi\hat{\mathbf{n}},\chi),
\label{eq:kapp1}
\end{eqnarray}
where the lensing kernel $W^{\kappa}(\chi)$ for spatially-flat Universe is~\citep{Schneider98,Waerbeke14}:
\begin{eqnarray}
W^{\kappa}=\frac{3}{2}\left(\frac{H_{0}}{c}\right)^{2}\Omega_{\rm
m}g(\chi)\frac{\chi}{a}.
\end{eqnarray}
\begin{eqnarray}
g(\chi)=\int^{\infty}_{\chi} \der \chi'n_{\rm S}(\chi')\left(\frac{\chi'-\chi}{\chi'} \right), \label{eq:g-chi}
\end{eqnarray}
where $n_{\rm S}(\chi)$ is the distribution of source as a function of comoving distance. To be consistent with RCSLenS source distribution, we use the fitting formulae as given in eq.~(12) of~\citet{Hojjati17}.

The Fourier transform of 2-D lensing convergence (Eq.~(\ref{eq:kapp1})) is
\begin{eqnarray}
\kappa_{\ell}(M,z) &=& \delta_{\rm 2D}(\ell; M,z) \nonumber \\
&\simeq &
\frac{W^{\kappa}}{\chi^{2}}\tilde{\delta}_{\rm
3D}\left(\frac{\ell}{\chi(z)};M,z \right) \nonumber \\
& = & \frac{W^{\kappa}(z)}{\chi^{2}(z)}\frac{1}{\rho_{\rm m}(z)} \nonumber \\
& \times &\int \der r (4\pi r^{2})
\left(\frac{\sin(\ell r/\chi)}{\ell r/\chi}
\right)\rho(ar;M,z),
\label{eq:kell1}
\end{eqnarray}
where $r$ is a comoving radius ($ar$ is the physical radius). $\rho(ar;M,z)$ is dark matter density profile which we use the NFW profile~\citep{Navarro97}. The $\gamma_{\rm t}(\ell| M,z)$ is strongly related to $\kappa_{\ell}(M,z)$ through a rotation (see Eq.~(\ref{eq:gamma-t-theta}) in Sec.~\ref{sec:angular}), and the resultant correlation function differs by the Bessel function. So we will need the $\kappa_{\ell}(M,z)$ to compute $C^{y-\gamma_{\rm t}}_{\ell}$ and then $\xi^{y-\gamma_{\rm t}}(\theta)$.

To facilitate the calculation of Eq.~(\ref{eq:kell1}), we define $x=a(z)r/r_{\rm s}$. $r_{\rm s}$ is a characteristic scale
radius of the profile in physical unit. Then by defining
$\ell_{\rm s}=a\chi/r_{\rm s}$ (a characteristic multipole
moment), one can transform Eq.~(\ref{eq:kell1}) into (see also
\cite{Hill13})
\begin{eqnarray}
\kappa_{\ell}(M,z)
&=& \frac{4\pi r_{\rm s}}{a\ell^{2}_{\rm
s}}\frac{W^{\kappa}}{\rho_{\rm m}(z)} \nonumber \\
& \times &
\int^{\infty}_{0} \der x x^{2}\left(\frac{\sin(\ell x/ \ell_{\rm s})}{\ell
x/\ell_{\rm s}} \right) \rho\left(xr_{\rm s};M,z\right), \nonumber \\ \label{eq:yell1}
\end{eqnarray}
where the scale radius $r_{\rm s}$ can be calculated from $r_{\rm s}=r_{\rm vir}/c$. $c$ is the concentration parameter which we use the simulation result from~\citet{Duffy08} as
\begin{eqnarray}
c=\frac{5.72}{(1+z)^{0.71}}\left(\frac{M}{10^{14}h^{-1}
\msun} \right)^{-0.081}. \label{eq:duffy-c}
\end{eqnarray}

The relation between virial mass and radius is
\begin{eqnarray}
M=\frac{4\pi}{3}\left[\Delta_{\rm c}(z)\rho_{\rm c}(z)
\right]r^{3}_{\rm vir}, \label{eq:Mvir}
\end{eqnarray}
where
\begin{eqnarray}
\Delta_{\rm c}(z)=18\pi^{2}+82[\Omega(z)-1]-39 [\Omega(z)-1]^{2},
\label{eq:Delta-c}
\end{eqnarray}
is the density contrast of virialized halo. $\Omega(z)=\Omega_{\rm m}(1+z)^{3}/E^{2}(z)$ ($\Omega_{\rm
m}$ is the current fractional matter density,
$E^{2}(z)=\Omega_{\rm m}(1+z)^{3}+\Omega_{\Lambda}$). $\rho_{\rm
c}(z)$ is the critical density at redshift $z$, which is
$\rho_{\rm c}(z)=(3H^{2}_{0}/8\pi G)E^{2}(z)$. Therefore, the
dependence of $r_{\rm vir}$ on virial mass $M$ is
\begin{eqnarray}
\left(\frac{r_{\rm vir}}{1\,h^{-1}\textrm{Mpc}} \right) &=& 9.51\times \left(\frac{M/(10^{15}\,h^{-1}\msun)}{\Delta_{\rm
c}(z)E^{2}(z)} \right)^{1/3}. 
\label{eq:rvir}
\end{eqnarray}
Therefore, through Eqs.~(\ref{eq:Delta-c}) and (\ref{eq:rvir}), $r_{\rm s}$ is a function of virial mass and redshift. To calculate the UPP model, one needs to calculate the value of $R_{500}$ (Eq.~(\ref{eq:unipres})) given $(M,z)$. One can use the formula in appendix B of~\citet{Planck-dispersion} to calculate $M_{500}$ and then $R_{500}$. Thus the entire $y_{\ell}$ function depends on virial mass and redshift.

\subsection{The power spectra of halo model}

Putting together the Fourier transform of lensing kernel and Compton-$y$ parameter, the 1-halo, 2-halo term of Compton-$y$ parameter and lensing convergence field $\kappa$ can be calculated as~\citep{Ma15}
\begin{eqnarray}
C^{y\kappa,1\rm h}_{\ell} & = & \int^{z_{\rm max}}_{0} \der
z\frac{\der^{2}V}{\der z\der \Omega}\int^{M_{\rm max}}_{M_{\rm min}} \der M \left(\frac{\der
n}{\der M}
\right) \nonumber \\
& \times & (y_{\ell}(M,z)\kappa_{\ell}(M,z)) \label{eq:Clyk1h} \\
C^{y\kappa,2\rm h}_{\ell} & = & \int^{z_{\rm max}}_{0} \der z \frac{\der^{2}
V}{\der z \der \Omega} P^{\rm lin}_{\rm m}\left(k=\frac{\ell +1/2}{\chi(z)},z\right) \nonumber \\
& \times &
\left[\int \der M \left(\frac{\der n}{\der M}
\right) b(M,z) y_{\ell}(M,z) \right] \nonumber \\
& \times & \left[\int \der M
\left(\frac{\der n}{\der M} \right) b(M,z) \kappa_{\ell}(M,z)
\right],
\label{eq:Clyk2h}
\end{eqnarray}
where $\der^{2}V/\der z \der \Omega= c\chi^{2}/H(z)$ is the comoving
volume element per redshift per steradians. $\der n/\der M$ and $b(M,z)$ are the halo mass function and gravitational bias function, which we use the Sheth \& Tormen function respectively~\citep{sheth1999}. The total correlation function between $y$ and $\kappa$ is
\begin{eqnarray}
C^{y\kappa}_{\ell}=C^{y\kappa, 1\rm h}_{\ell}+ C^{y\kappa, 2\rm h}_{\ell}.
\end{eqnarray}

The angular correlation function $\xi^{y-\kappa}(\theta)$ and $\xi^{y-\gamma_{\rm t}}(\theta)$ with the flat-sky approximation are obtained via
\begin{eqnarray}
\xi^{y-\kappa}(\theta) &=& \frac{1}{2\pi}\sum_{\ell}\ell C^{\kappa y}_{\ell}J_{0}(\ell \theta)B^{y}_{\ell}B^{\kappa}_{\ell} , \nonumber \\
\xi^{y-\gamma_{\rm t}}(\theta) &=& \frac{1}{2\pi} \sum_{\ell}\ell C^{\kappa y}_{\ell}J_{2}(\ell \theta)B^{y}_{\ell},
\end{eqnarray}
where the beam function for $y$-map and $\kappa$-map are
\begin{equation}
B^{\kappa,y}_{\ell}=\exp\left(-\ell^{2}\sigma^{2}_{\rm b}/2 \right),
\end{equation}
where $\sigma_{\rm b}=\theta_{\rm FWHM}/\sqrt{8\ln 2}=0.00742\left(\theta_{\rm FWHM}/1^{\circ} \right)=1.237 \times 10^{-3}$, for the $\theta_{\rm FWHM}=10\,$arcmin of both {\it Planck} $y-$map and lensing convergence map ($\kappa$). For shear map, it was {\it not} smoothed ($\theta_{\rm FWHM}=0.0$), so there is no beam function associated with it. The detail calculation is presented in Appendix~\ref{sec:angular}.

\subsection{Likelihood analysis}

Finally we formulate the $\chi^{2}$ for both $\xi^{y-\kappa}(\theta)$ and $\xi^{y-\gamma_{\rm t}}(\theta)$ as
\begin{eqnarray}
\chi^{2}(\mathbf{p}) &=& \sum^{N_{\rm bin}}_{i=1}\sum^{N_{\rm bin}}_{j=1}\left(\xi^{\rm data}(\theta_{i})- \xi^{\rm the}(\theta_{i}|\mathbf{p}) \right)\left(C^{-1} \right)_{ij} \nonumber \\
& \times & \left(\xi^{\rm data}(\theta_{j})- \xi^{\rm the}(\theta_{j}|\mathbf{p}) \right),
\end{eqnarray}
in which $\mathbf{p}=(P_{0},c_{500},\alpha,\beta)$ denotes the pressure profile parameter set. We calculate $\chi^{2}_{\xi^{y-\kappa}}$ and $\chi^{2}_{\xi^{y-\gamma_{\rm t}}}$ respectively and then formulate the maximum likelihood as $\mathcal{L}\sim \exp(-\chi^{2}/2)$.

To calculate the posterior distribution of the UPP parameters, we utilize the Markov Chain Monte Carlo (MCMC) method to populate the samplings in the parameter space. We use the Metropolis-Hastings algorithm to determine the probability of accepting the new points in the chain~\citep{Metropolis53,Hastings70}. The density matrix proposed is obtained by Gaussian sampler with adaptive step size~\citep{Doran04,Gong19}. We set uniform flat prior for the UPP parameters in the following range: $P_{0}\in(0,20)$, $c_{500} \in (0,10)$, $\alpha \in (0,10)$, and $\beta \in (0, 10)$. We run fifteen parallel chains in total, and obtain $10^{5}$ points for each chain after it reaches the convergence criterion with $R-1<0.01$~\citep{Gelman92}. We then perform the burn-in and thinning of the chains, and merge all chains into obtaining 1-D and 2-D posterior distribution of the UPP parameters.

\section{Results}
\label{sec:results}

\begin{figure*}
\centerline{\includegraphics[width=5.0in]{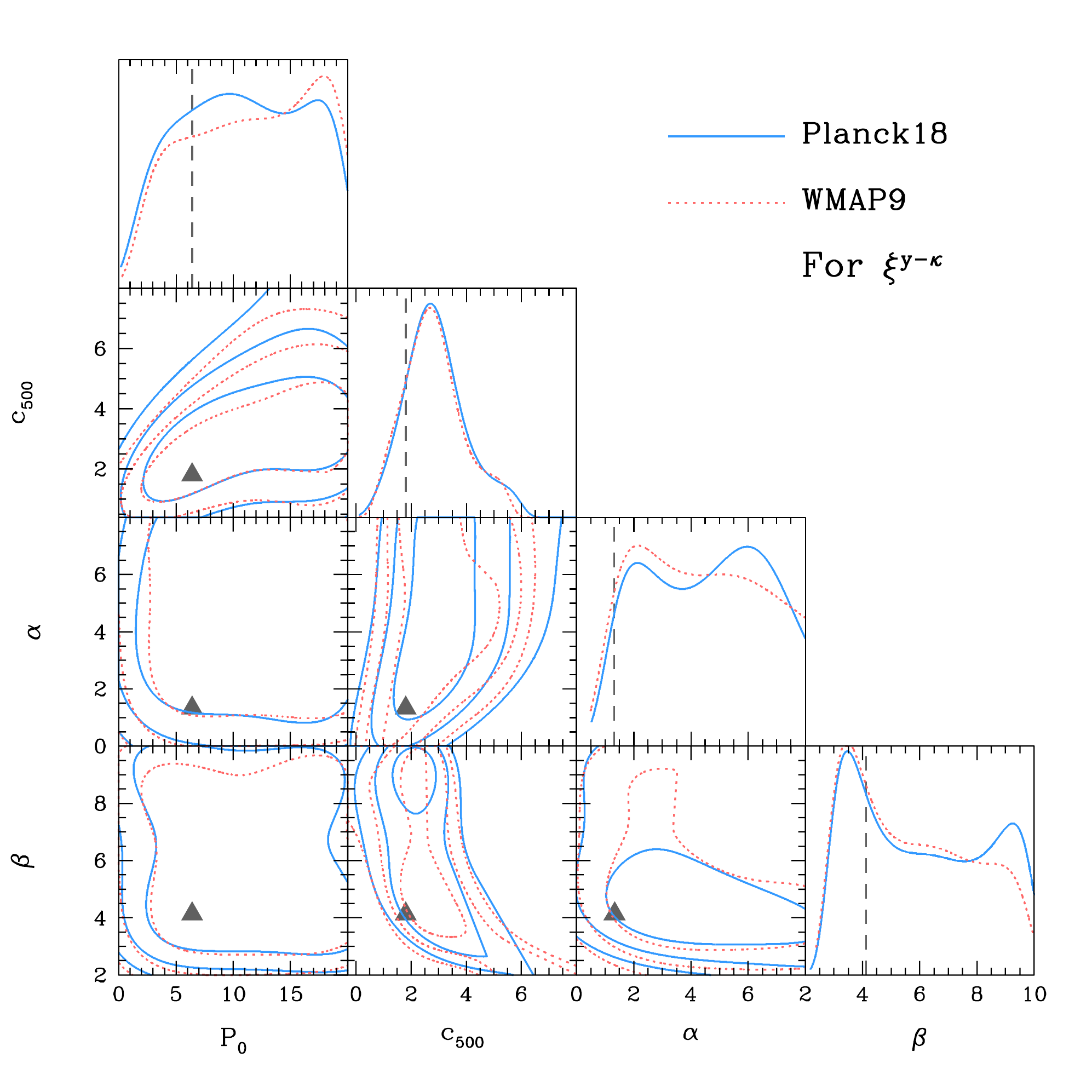}}
\caption{The joint constraints on UPP parameters ($P_{0}$, $c_{500}$, $\alpha$, $\beta$) with $\xi^{y-\kappa}$ correlation data by assuming {\it WMAP}-9 (red dashed contours) and {\it Planck}-2018 (blue solid contours) cosmological parameters. The contours show $68\%$, $95\%$ and $99\%$ C.L. respectively. The best-fitting UPP parameter values from~\citet{Planck-parameters18} are marked with grey triangles and dashed lines.} \label{fig:contour}
\end{figure*}

\begin{figure*}
\centerline{\includegraphics[width=5.0in]{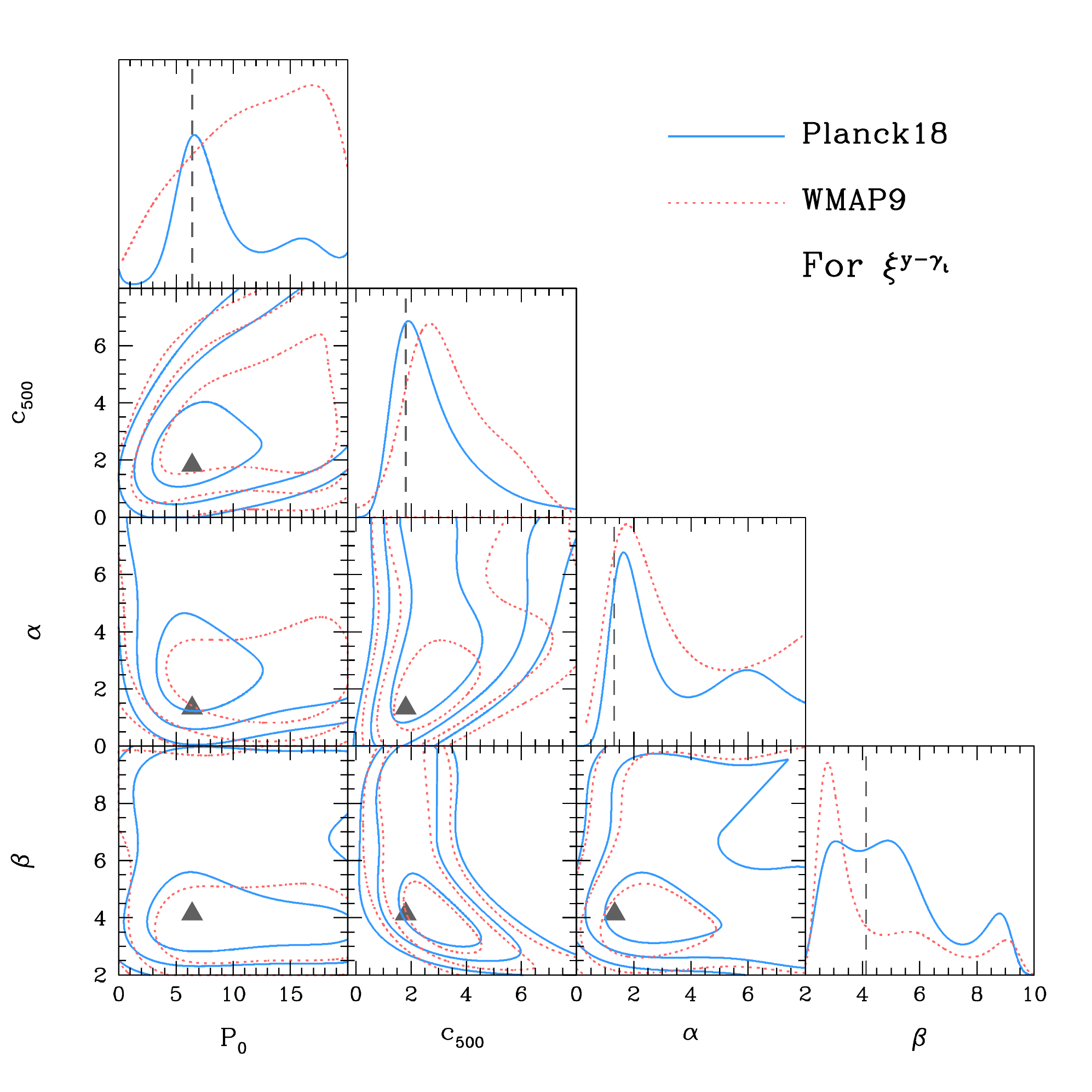}}
\caption{Same as Fig.~\ref{fig:contour} but for $\xi^{y-\gamma_{\rm t}}$ correlation data.} \label{fig:contour2}
\end{figure*}

\begin{figure*}
\centerline{\includegraphics[width=3.2in]{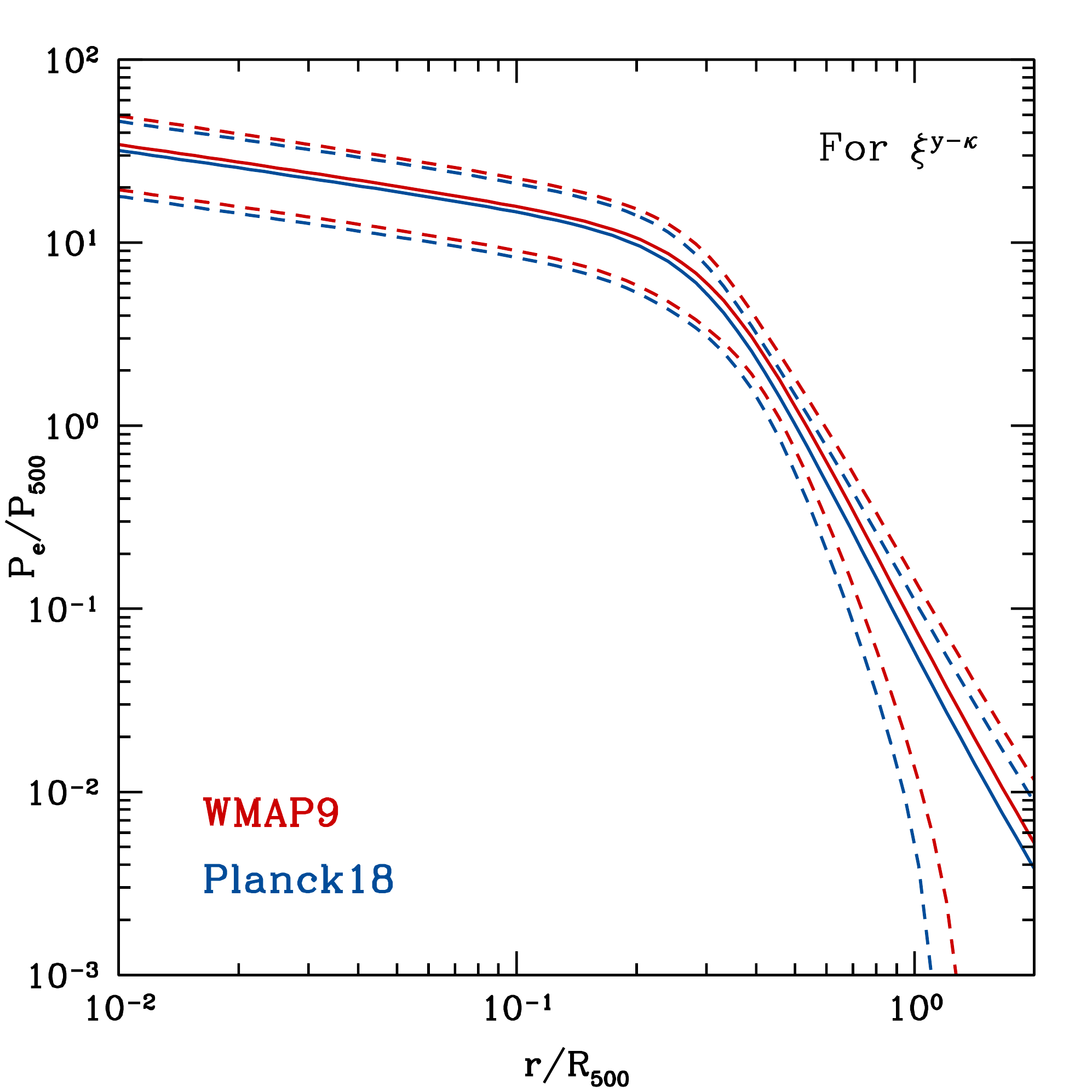}
\includegraphics[width=3.2in]{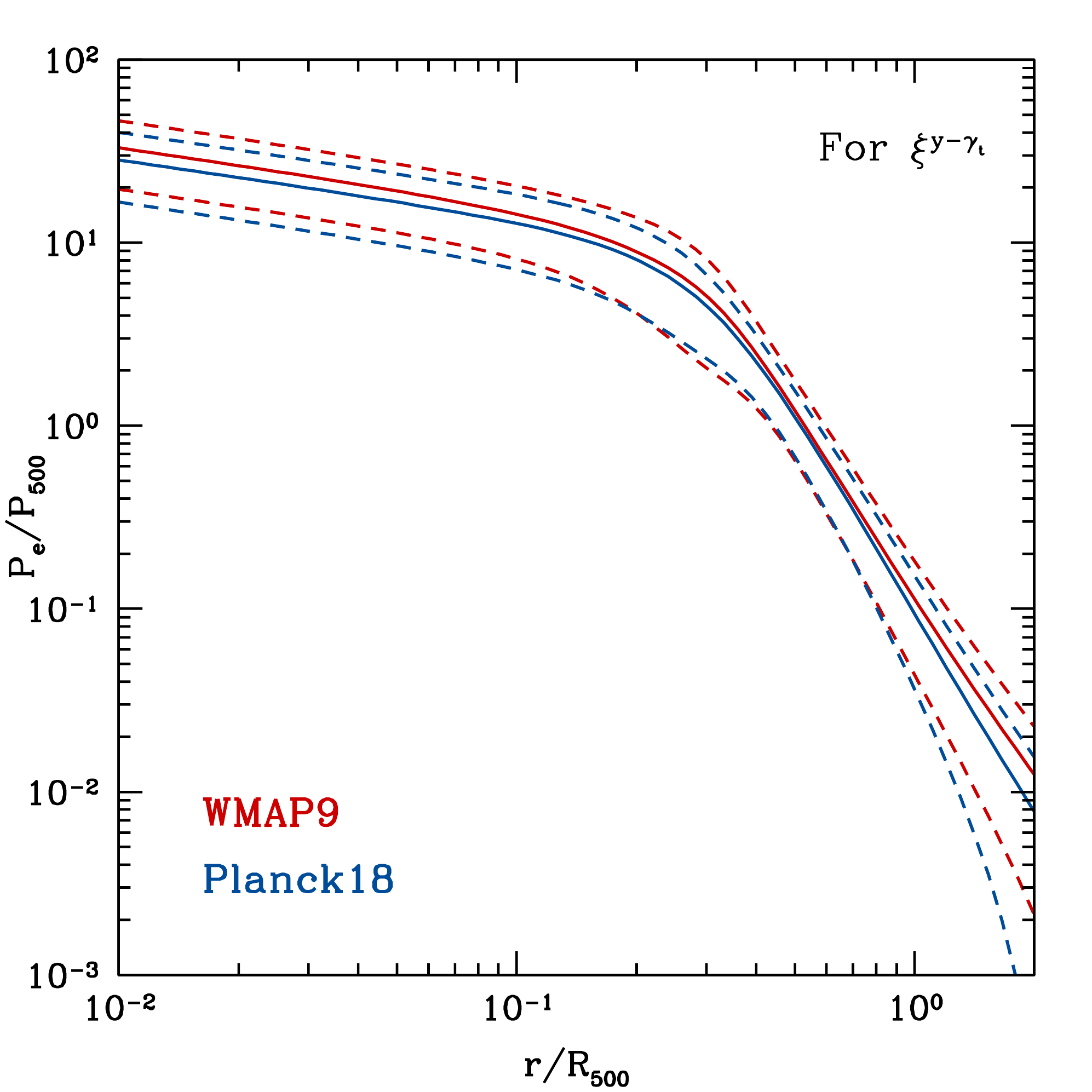}}
\caption{The mean value (solid lines) and standard deviation (dashed lines) of $P_{\rm e}/P_{500}$ (redshift independent which can be seen from Eqs.~(\ref{eq:unipres}) and (\ref{eq:P500}).) as a function of $r/R_{500}$ derived from the MCMC chains for {\it Planck}-2018 and {\it WMAP}-9 cosmologies. The left and right panels are for $\xi^{y-\kappa}$ and $\xi^{y-\gamma_{\rm t}}$ results. Here we assume $M_{500}=3 \times 10^{14}\msun$.} \label{fig:Pe}
\end{figure*}

\begin{figure*}
\centerline{\includegraphics[width=3.2in]{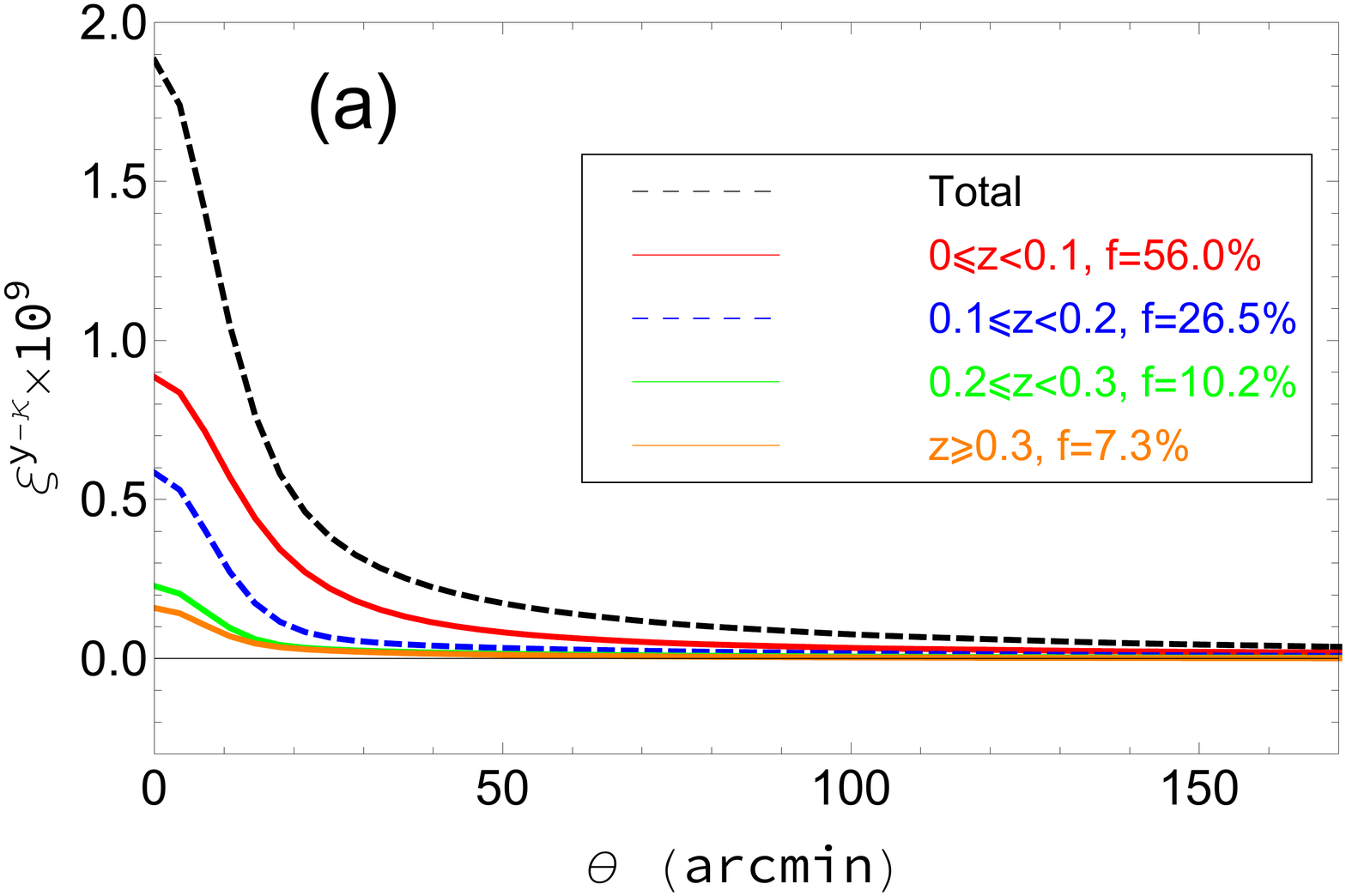}
\includegraphics[width=3.2in]{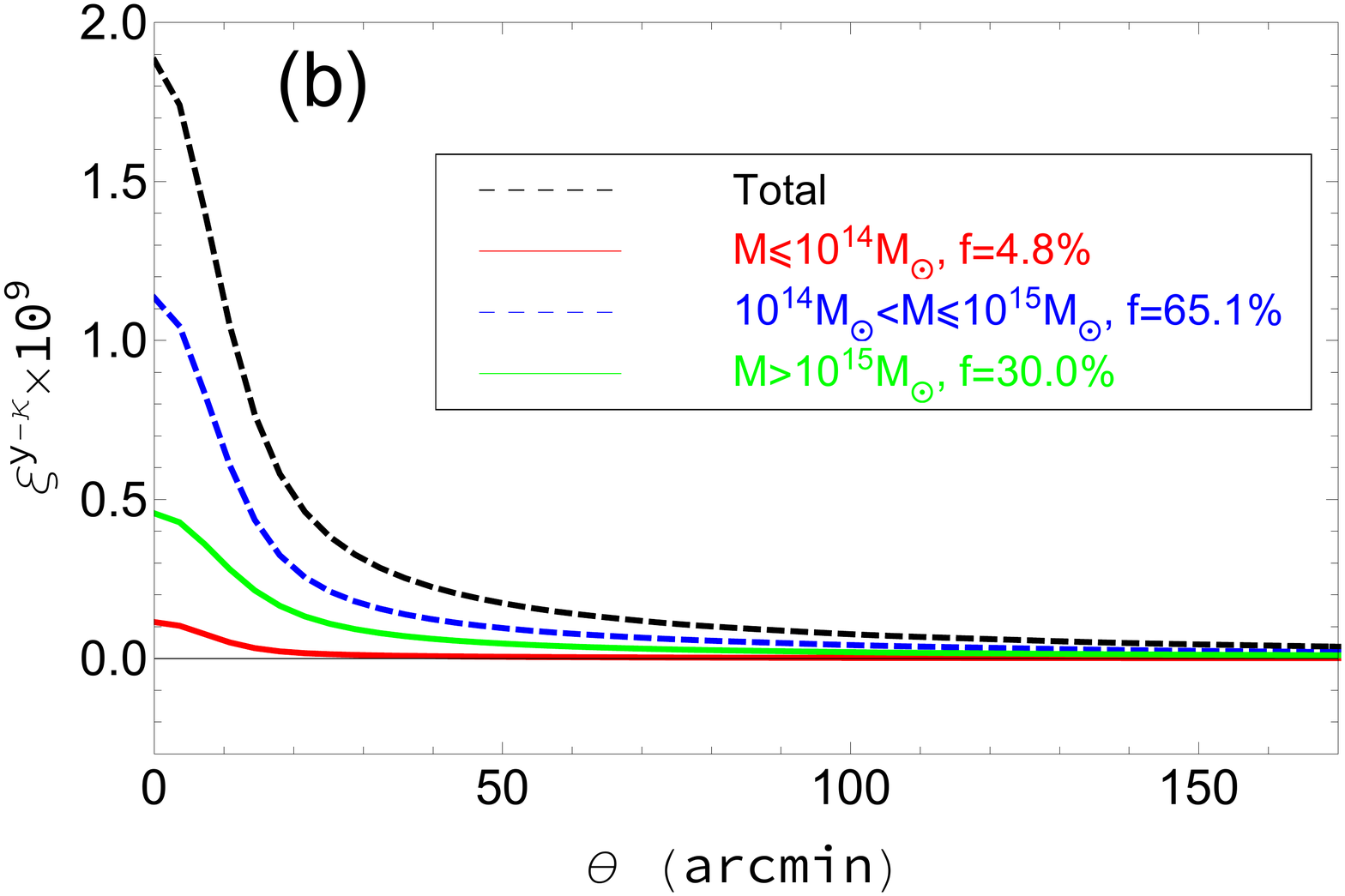}}
\centerline{\includegraphics[width=3.5in]{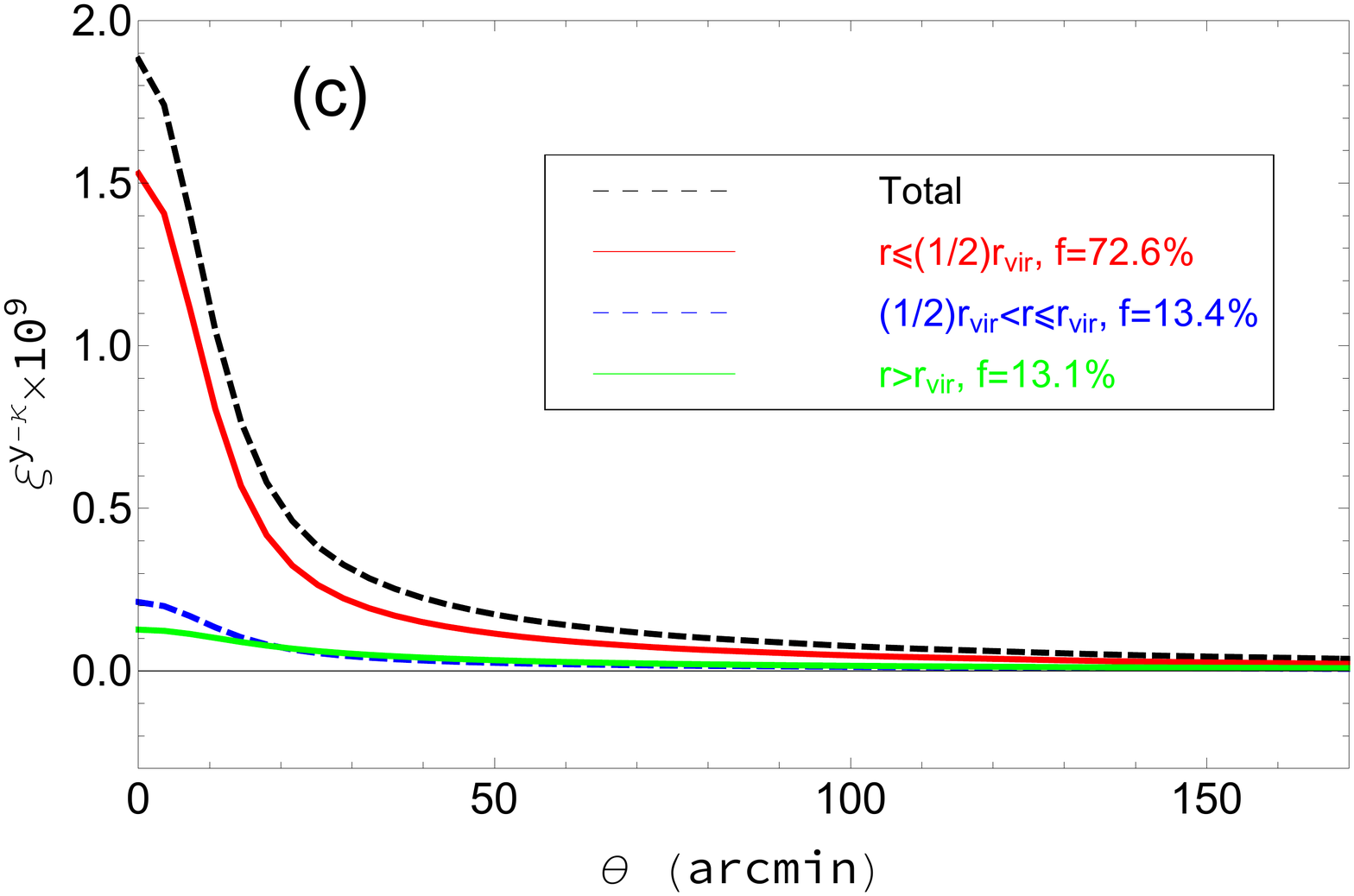}}
\caption{The separation of the $\xi^{y-\kappa}(\theta)$ signal in different redshift bins (panel (a)), mass bins (panel (b)) and radius bins (panel (c)). Lines with different colours and solid/dashed styles indicate the different redshift, mass and radius regimes. The $f$-value is the fractional contribution of each bin to the total signal curve.} \label{fig:bin}
\end{figure*}

\subsection{The UPP parameters}
In the halo model, the underlying cosmological parameters affect the cross-correlation function. For example, values of $\Omega_{\rm m}h^{2}$, $\sigma_{8}$ and $n_{\rm s}$ have an impact on the amplitude and the shape of the matter power spectrum. But in our study, we cannot release all of these cosmological parameters because the likelihood will be too loose and unable to constrain any parameter in a realistic range. Besides, as one can see from Table~\ref{tab:cosmology}, the difference between parameter values from {\it Planck} and {\it WMAP} are not big enough to predict to be very different. We therefore fix the cosmological parameter values to be {\it Planck}-2018~\citep{Planck-parameters18} and {\it WMAP} 9-year best-fitting cosmological parameter values~\citep{Hinshaw13}, and list them in Table~\ref{tab:cosmology}. We will compare the results of constraints for these two settings of background cosmology.

We release four parameters $(P_{0},c_{500},\alpha,\beta)$ in the likelihood chain. As shown in Eq.~(\ref{eq:px}), $P_{0}$ controls the amplitude of UPP, $c_{500}$ is the pressure concentration which controls the relation between $R_{500}$ and $R_{\rm s}$, ($\alpha$, $\beta$, $\gamma$) control the slope of the profile. If we release all five parameters, the pressure profile can take any amplitude and shape so the constraints will be very loose. Therefore we fix the inner slope parameter $\gamma=0.31$, because its value is quite precisely determined by {\it XMM-Newton} observations of the cluster profile~\citep{Planck13-pressure}. We release ($P_{0},c_{500},\alpha,\beta$) in the likelihood chain. 
  
In Figs.~\ref{fig:contour} and~\ref{fig:contour2}, we plot the joint constraints on the UPP parameter set from $\xi^{y-\kappa}$ and $\xi^{y-\gamma_{\rm t}}$ data sets, by assuming {\it Planck} and {\it WMAP} cosmological parameters as blue solid line and red dashed line. The values obtained are shown in Table~\ref{tab-para}. We also plot the best-fitting values of the UPP parameters from~\citet{Planck13-pressure} as grey triangles and dashed vertical lines, which was obtained by stacking a sample of 62 massive clusters in mass range $2 \times 10^{14}\msun< M_{500} < 2 \times 10^{15} \msun$. One can see that the current constraints from the correlation function for both assumed cosmologies are consistent with the values obtained from {\it Planck} 62-clusters within $1\sigma$ C.L. The marginalized likelihood of $c_{500}$, $\alpha$ and $\beta$ are quite consistent between {\it WMAP} and {\it Planck} assumed cosmologies. For {\it WMAP}-9 cosmology and $\xi^{\gamma_{\rm t}-y}$ estimator we find $c_{500} = 2.68^{+1.46}_{-0.96}$  and $\alpha=1.75^{+1.29}_{-0.77}$ (shape index for intermediate radius); for {\it Planck}-2018 cosmology we find $c_{500} = 1.91^{+1.07}_{-0.65}$ and $\alpha = 1.65^{+0.74}_{-0.5}$. Only for $P_{0}$ the assumed {\it WMAP} cosmology of $\xi^{y-\gamma_{\rm t}}$ prefers a higher value, but it cannot exclude the lower value preferred by {\it Planck} cosmology case due to its broad distribution. The reason that {\it WMAP} cosmology parameter case prefers a slightly higher value of $P_{0}$ is because its best-fitting $\sigma_{8}$ value is lower than {\it Planck} value. Therefore, to balance out the decrease amplitude of $\kappa_{\ell}$ in Eq.~(\ref{eq:Clyk1h}) one needs to have a slightly higher amplitude of $y_{\ell}$ which boosts up the $P_{0}$ value.

In Table~\ref{tab-para}, we list the numerical values of the fitting results. For comparison, we also list the parameters in~\citet{Arnaud10} obtained from stacking $33$ local ($z<0.2$) clusters with mass $10^{14}\msun<M_{500}<10^{15}\msun$; the parameters in~\citet{Planck13-pressure} obtained from stacking $62$ nearby ($z<0.5$) clusters with mass $2\times 10^{14}\msun<M_{500}<2 \times 10^{15}\msun$; the parameters in~\citet{Gong19} obtained from stacking $101,407$ central, nearby galaxies ($z<0.5$) with mass $10^{13}\msun<M_{500}<10^{15}\msun$. One can see that for these four parameters, our current results from {\it Planck} and {\it WMAP} cosmologies are consistent with \citet{Planck13-pressure} within $1\sigma$ C.L. The $\beta$-value shows $~2\sigma$ deviation from~\citet{Arnaud10} but the error-bar is still too big to derive a definite answer. The $\chi^{2}_{\rm min}$ is shown in the last column in our fitting. The values for {\it WMAP} and {\it Planck} cosmologies are close to each other. The degree of freedom in this fitting is $8-4=4$, so $\chi^{2}_{\rm min}/N_{\rm dof}=0.71$ and $0.68$ for {\it WMAP}-9 and {\it Planck} cosmology $\xi^{y-\kappa}$ data set, and $1.64$ for both {\it WMAP}-9 and {\it Planck} cosmology $\xi^{y-\gamma_{\rm t}}$ data set. Therefore, the data is very well fitted by the prediction from the halo model, which can also be seen in Fig.~\ref{fig:Planck-WMAP}. The Fig.~\ref{fig:Planck-WMAP} shows the excellent consistency between the prediction of the halo model with the data from RCSLenS data cross-correlation with {\it Planck} $y-$map. The bottom panel shows the difference between {\it WMAP} best-fitting curve and that of {\it Planck}, together with the residual of the data (the data minus the {\it Planck} prediction). One can see that the current data is unable to distinguish the small difference in {\it WMAP} and {\it Planck} prediction. Future data with tightened up constraints can achieve the distinction.

To show the current best-fitting UPP, we plot the $P_{\rm e}/P_{500}$ value as a function of $r/R_{500}$ for a fixed mass $M_{500}=3\times 10^{14}\msun$ in Fig.~\ref{fig:Pe}. This $P_{\rm e}/P_{500}=F(M_{500})\mathbb{P}(x)$ is independent of the redshift evolution factor as seen from Eqs.~(\ref{eq:unipres}) and (\ref{eq:P500}). We run through the constrained parameter spaces of {\it Planck} and {\it WMAP} cosmologies, and plot the best-fitting function $P_{\rm e}/P_{500}$, and its $\pm 1\sigma$ boundary lines. One can see that although the errors of the UPP parameters are still quite large, the shape of the UPP is relatively fixed, and the amplitude of the profile does not vary by a factor of $3$. Our result is, to date, the most precise determination of the pressure profile of galaxy clusters if allowing the four parameters of UPP to vary\footnote{Previous works tried to constrain UPP parameters to various extent. For example, \citet{Romero15} used MUSTANG and Bolocam data to constrain the $\gamma$ parameters for Abell 1835 and MACS0647, while fixing all other parameters. \citet{Romero17} fitted 14 clusters' UPP profile and obtained tight constraints but only allowing $c_{500}$, $\gamma$ and $P_{0}$ to vary. \citet{Sayers13} used Bolocam (Caltech) observation of a set of 45 massive clusters' images and constrained the UPP parameter but only reported the best-fitting values without confidence levels (full likelihood).}. The universal pressure profile under {\it WMAP} and {\it Planck} cosmological parameters are also consistent with each other at all radii out to $R_{500}$.

\subsection{Contributions from different mass, radius and redshift bins}

Given the current constraints on the UPP parameters, we further investigate the contribution of the signal of cross-correlation from baryons residing in different masses of halos, different redshift ranges and different radius relative to the centre of halos. Since both the real-space $\xi^{y-\kappa}$ and $\xi^{y-\gamma_{\rm t}}$ come from the Fourier transformation of power spectrum $C^{y-\kappa}_{\ell}$, here we only use the $\xi^{y-\kappa}(\theta)$ to investigate the fractional contribution. To investigate the fractional contribution from each redshift, mass and radius bin, we need to calculate the fraction of the area under the curve in each bin to the total area, i.e. 
\begin{eqnarray}
f=\frac{\int \der \theta \xi^{\rm bin}(\theta)}{\int \der \theta \xi^{\rm total}(\theta)}.
\end{eqnarray}

\subsubsection{Redshift bins}

To separate the signal into different redshift bins, we limit the integral in Eq.~(\ref{eq:Clyk1h}) into different redshift intervals. The results are shown in panel~(a) of Fig.~\ref{fig:bin}. The $f$-value quoted in the legend shows the contribution of signals in the respective bin relative to the total signal, by calculating the fraction of areas under the curve to the full curve.

One can see that most of the signal ($f=56.0\%$) of the correlation function $\xi^{y-\kappa}$ comes from the $z<0.1$ band. As the redshift increases, the contributions gradually decrease, and there is a small contribution ($f=7.3\%$) to the total signal from baryons that locate at $z>0.3$. This phenomenon suggests that the $\kappa-y$ cross-correlation is most sensitive to the low redshift regime of the ionized gas. As shown in~\citet{cenostriker2006} and~\citet{Bregman07}, most of the baryons in the warm-hot intergalactic medium (WHIM) phase with temperature in between $10^{5}$ to $10^{7}$\,K are at low redshifts ($z \leq 0.5$). Therefore, the cross-correlation signal is most sensitive to this WHIM gas. Besides, our result is quantitatively consistent with the result of redshift contribution found in~\citet{Hojjati15}. As one can see in the upper panel of fig.~2 of ~\citet{Hojjati15}, the major contribution of the cross-correlation signal is also from $z<0.3$, but the contribution from $z>0.3$ is slightly bigger than we obtained here. The reason is that \citet{Hojjati15} used the cosmo-OWLS simulation, which is an extension of the OverWhelmingly Large Simulations project~\citep{Schaye10}. The simulation was run with a significantly modified version of the Lagrangian TreePM-SPH code {\sc gadget3}~\citep{Springel05} which takes into account for the various AGN feedback models. Therefore, the source distribution and lensing kernel could be different in between this simulation code and RCSLenS data, but they are consistent with each other in a quantitative way generally.

\subsubsection{Mass bins}

To investigate the mass contribution from different regimes, we separate integral of the mass in Eq.~(\ref{eq:Clyk1h}) and the first square bracket in Eq.~(\ref{eq:Clyk2h}) to different intervals of the halo mass and plot their correlation function in panel (b) of Fig.~\ref{fig:bin}. One can see that the mass bin $10^{14}\msun<M_{\rm vir}<10^{15}\msun$ makes the major contribution ($f=65.1\%$) to the total signal. The halo mass regime of $M<10^{14}\msun$ contributes $4.8$ per cent of the correlation contribution, and $M>10^{15}\msun$ contributes $30.0$ per cent of the correlation contribution. This result shows that the majority of the $\kappa-y$ cross-correlation comes from halos with masses in between $10^{14}$--$10^{15}\msun$. The contribution from halos above or below this regime is small. This result is also consistent with the previous findings in~\citet{Hojjati15}. As shown in the second upper panel of fig.~2 in~\citet{Hojjati15}, if the mass cut of $M>5\times 10^{14}\msun$ is made to the simulation catalogue, then the correlation function drops down to $\sim 34$ per cent. This result is quantitatively consistent with our finding because the mass cut $M>10^{15}\msun$ reduces the correlation function to $\sim 28$ per cent level.

\subsubsection{Radius bin}

Then we calculate the fractional contribution of baryonic gas in different radius range to the centre of dark matter halo. We do this by separating the integral in Eq.~(\ref{eq:yell1}) into three different radius intervals, which corresponds to three ranges in $x$, i.e.
\begin{eqnarray}
0<r \leqslant r_{\rm vir}/2  \nonumber \\
r_{\rm vir}/2 < r \leqslant r_{\rm vir}  \nonumber \\
r>r_{\rm vir}.  \label{eq:r-interval}
\end{eqnarray}
Then we substitute their corresponding $y_{\ell}(M,z)$ value into Eqs.~(\ref{eq:Clyk1h}) and (\ref{eq:Clyk2h}) and calculate the correlation function and fractional contribution. We show our results in panel (c) of Fig.~\ref{fig:bin}. One can see that $72.6$ per cent of baryons contributed to the signal comes from the inner regime of the profile $r\leqslant r_{\rm vir}/2$ whereas outer regime $r>r_{\rm vir}$ contributes to about $13.1$ per cent. This phenomenon shows the contribution from the baryon located at larger radii is non-negligible, but it is comparably smaller than previous studies of hydro-simulation in~\citet{Hojjati15}. In the third upper panel and table~2 of~\citet{Hojjati15}, one can see that the outer radius regime contributes to roughly $11\%+24\%= 35\%$ of the total signal, which is larger than the result we obtain here. Nonetheless, qualitatively, both studies show that the major contribution of the correlation signal comes from the low-radius part of the gaseous halo profile.

\section{Conclusion}

In this paper, we make use of the cross-correlation between the convergence and shear measurements of the weak gravitational lensing from RCS lensing survey (RCSLenS) with the thermal Sunyaev-Zeldovich signal measured by {\it Planck} satellite to study the universal pressure profile (UPP) of galaxy clusters. The cross-correlation signals of $\xi^{\kappa-y}(\theta)$ and $\xi^{\gamma_{\rm t}-y}(\theta)$ are measured at $7.1\sigma$ C.L. and $8.1\sigma$ C.L. individually (including sample variance), and we combine these two data sets to probe the cluster pressure profile.

We first derive the theoretical cross-correlation signal $\xi^{\kappa-y}(\theta)$ and $\xi^{\gamma_{\rm t}-y}(\theta)$ by using the halo model. In this derivation, we take into account the source distribution of RCSLenS data and the {\it Planck} beam in the estimates. We first compute the 1-halo and 2-halo terms of the cross-correlated power spectra and transform them into angular space to calculate the correlation function for both $\kappa-y$ and $\gamma_{\rm t}-y$ correlation. We adopt both {\it WMAP} 9-year and {\it Planck} 2018 best-fitting cosmological parameters in the calculation. We then employ the MCMC technique to sample the parameters space of UPP parameters $(P_{0},\alpha,\beta,c_{500})$ and calculate the posterior distributions. 

We present our results of numerical fitting in Table~\ref{tab-para}. Except for $P_{0}$, the estimates of $(\alpha,\beta,c_{500})$ are consistent in between {\it Planck} and {\it WMAP} cosmologies. The difference in $P_{0}$ (amplitude of UPP) is due to the difference in best-fitting $\Omega_{\rm m}$ values between {\it Planck} and {\it WMAP} cosmologies, but they are broadly consistent with each other due to the large error bars. The {\it Planck} measurement for UPP parameters is $(P_{0},c_{500},\alpha,\beta)=(9.68^{+10.02}_{-7.11}, 2.71^{+0.92}_{-0.93}, 5.97^{+1.81}_{-4.73}, 3.47^{+1.39}_{-0.60})$ for $\xi^{y-\kappa}$ fitting and $(6.62^{+2.06}_{-1.65}, 1.91^{+1.07}_{-0.65}, 1.65^{+0.74}_{-0.50}, 4.88^{+1.18}_{-2.46})$ for $\xi^{y-\gamma_{\rm t}}$ fitting, for which the UPP is constrained to be within a factor of $3$ in shape and magnitude. We then separate the best-fitting halo models into different redshift, halo mass and radii regime. We find that most of the baryons that contribute to the cross-correlation signals are located at low redshift $z<0.1$, halo mass in between $10^{14}$--$10^{15}\msun$ and within half of the virial radius ($r\leqslant r_{\rm vir}/2$). But the baryons associated with low mass halo, and those are diffused outside the virial radii also have non-negligible contributions to the total signal. 

Our results are consistent with the previous findings in~\citet{Ma15} and~\citet{Hojjati15,Hojjati17} that there is a non-negligible signal coming from the baryons gravitationally associated with low-mass halos and diffuse outside the halo virial radius. We can also compare our best-fitting halo model prediction (Fig.~\ref{fig:Planck-WMAP}) with the hydrodynamic simulation shown in~\citet{Hojjati15}. One can see that the hydrodynamic simulation with AGN 8.0 feedback model is the closet match to our prediction and the data. This corresponds to the temperature by which neighbouring gas is raised due to feedback is around $\Delta T_{\rm heat}\simeq 10^{8.0}\,{\rm K}$. Increasing the value of $\Delta T_{\rm heat}$ results in much stronger feedback events and also more bursty feedback. One should notice that recent study with clustering and weak lensing correlation from Dark Energy Survey (DES) excluded $\Delta T_{\rm heat}\simeq 10^{8.7}\,{\rm K}$ at $2\sigma$ C.L.~\citep{Huang20}.

In conclusion, we should be clear that the majority of the cross-correlation signal between Compton-$y$ map and weak gravitational lensing comes from high-mass halos and central region, but the baryons that diffuse outside the virial radius and associate with low-mass halos contribute non-negligibly. Our Markov-Chain Monte-Carlo likelihood shows that the halo concentration is consistent with the previous finding by using cluster studies~\citep{Planck13-pressure}. In addition, the 2-pt correlation function from the halo model can fully describe the $\kappa$-$y$ cross-correlation. Future more precise measurement on this cross-correlation, and with better modelling of the pressure profiles of clusters and filaments has the potential to determine the baryonic distribution ultimately.

\section*{Data Availability}
The Planck data used in this paper is a public dataset that can be downloaded from Planck Legacy Archive (\url{https://pla.esac.esa.int}). The RCSLenS data is available at the webpage \url{http://www.rcslens.org/}. The cross-correlation data and its covariance matrix are not public datasets, which can be requested by contacting the authors.

\section*{Acknowledgements}
YZM acknowledges the support of NRF-120385, NRF-120378, and NSFC-11828301. YG acknowledges the support of NSFC-11822305, NSFC-11773031, NSFC-11633004, the Chinese Academy of Sciences (CAS) Strategic Priority Research Program XDA15020200, the NSFC-ISF joint research program No. 11761141012, and CAS Interdisciplinary Innovation Team. TT acknowledges funding from the European Union's Horizon 2020 research and innovation programme under the Marie Sk{\l}odowska-Curie grant agreement No 797794.

\bibliographystyle{mnras}
\bibliography{ref} 

\appendix
\section{Angular correlation function}
\label{sec:angular}
\begin{figure*}
\centerline{\includegraphics[width=3.3in]{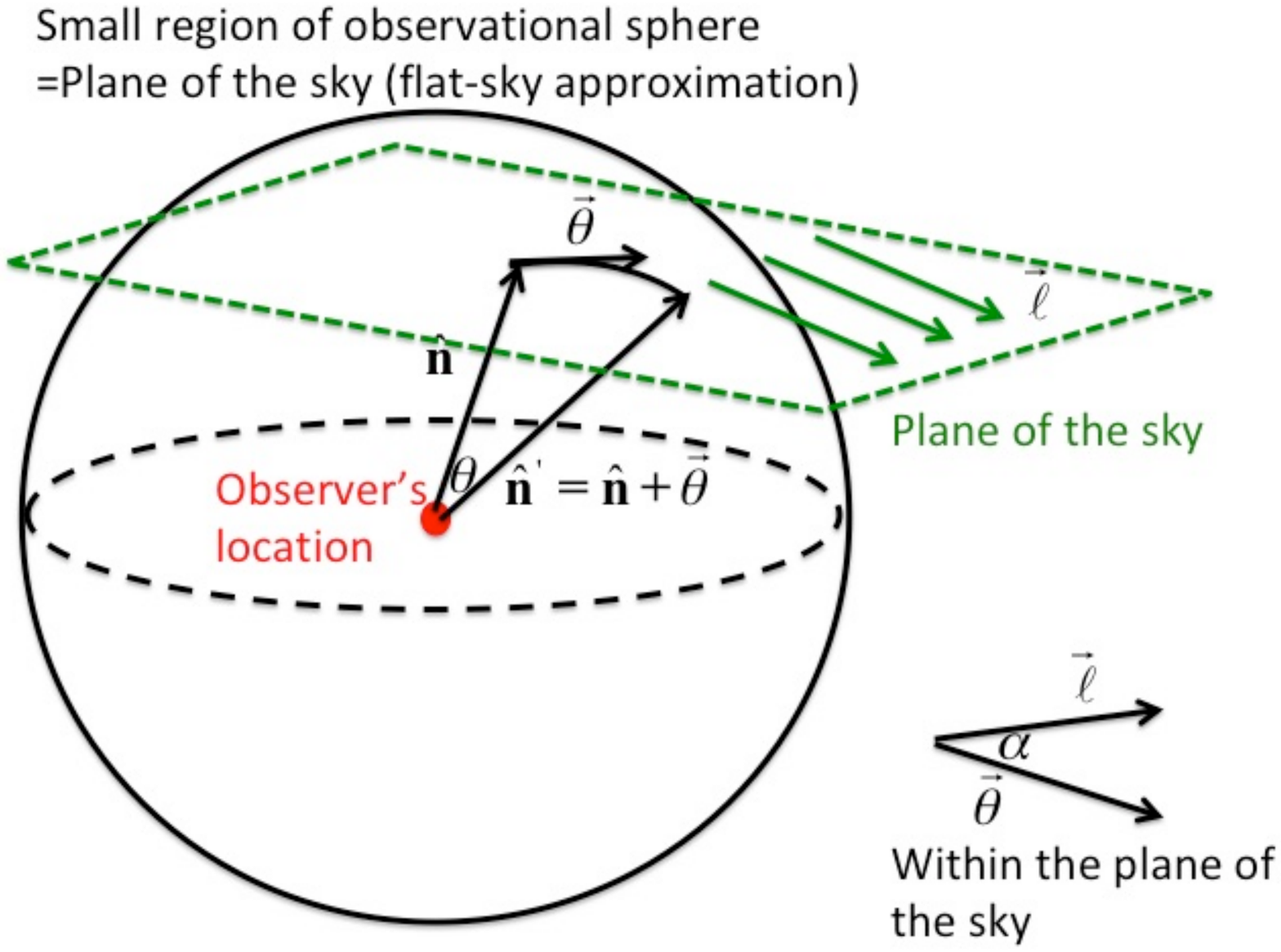}
\includegraphics[width=4.3in]{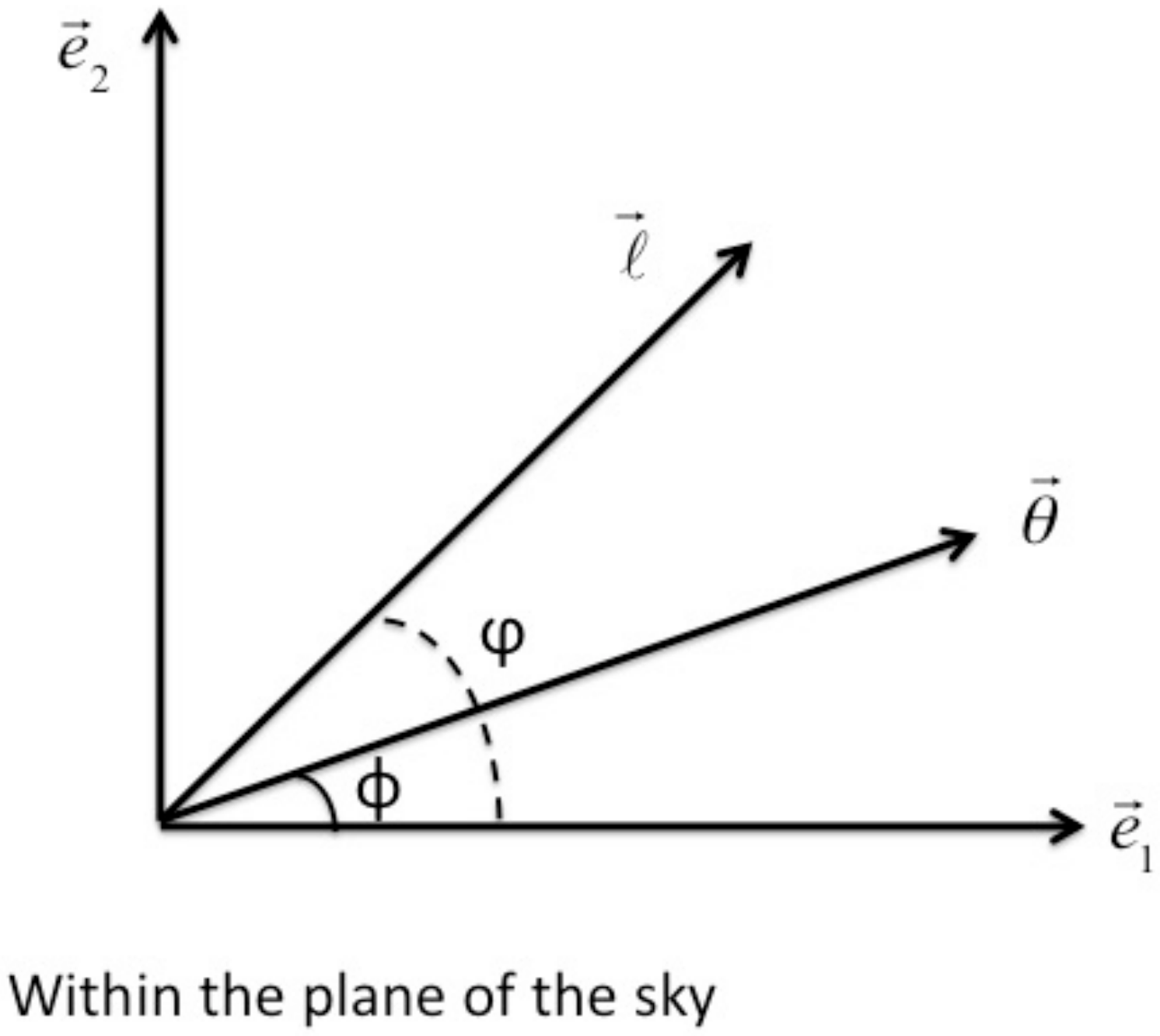}}
\caption{The geometric relation between $\vec{\ell}$, $\vec{\theta}$ within the plane of the sky with coordinate \{${\vec{\mathbf{e}}}_{1}$, ${\vec{\mathbf{e}}}_{2}$\}.} \label{fig:angle}
\end{figure*}

Here we lay out the detail calculation of the cross-correlation functions of $\kappa$ and $y$, and $\gamma_{\rm t}$ and $y$, by using {\it flat-sky} approximation~\footnote{If one uses a full-sky formalism, the sum over modes should be over Wigner-D matrix~\citep{Kilbinger17}.}. Fig.~\ref{fig:angle} helps to understand the geometric relations between different angles and quantities.

\subsection{$\xi^{\kappa y}(\theta)$}
Both $y(\hat{\mathbf{n}})$ and $\kappa(\hat{\mathbf{n}})$ are the random fields on the sky, so they are functions of sky position $\hat{\mathbf{n}}$. $\vec{\mathbf{\theta}}$ is the angular separation between the selected two points of the two fields (Fig.~\ref{fig:angle}). The correlation function can be written as
\begin{eqnarray}
\xi^{\kappa y}(\theta) &=& \int \frac{\der^{2}\hat{\mathbf{n}}}{4 \pi} \langle y(\hat{\mathbf{n}})\kappa(\hat{\mathbf{n}}+\vec{\theta}) \rangle \nonumber \\
&=& \int \frac{\der^{2}\hat{\mathbf{n}}}{4 \pi} \left\langle \left[\int\frac{\der^{2}\vec{\ell}}{(2\pi)^{2}}y(\vec{\ell}){\rm e}^{-i \vec{\ell}\cdot \hat{\mathbf{n}}} \right]  \right. \nonumber \\ 
& \times & \left.\left[\int\frac{\der^{2}\vec{\ell}'}{(2\pi)^{2}}\kappa(\vec{\ell}'){\rm e}^{i \vec{\ell}'\cdot (\hat{\mathbf{n}}+\vec{\theta})} \right]
 \right\rangle \nonumber \\
&=& \int\frac{\der^{2}\hat{\mathbf{n}}}{4\pi}\int\frac{\der^{2}\vec{\ell}}{(2\pi)^{2}} \frac{\der^{2}\vec{\ell}'}{(2\pi)^{2}} \langle y(\vec{\ell})\kappa(\vec{\ell}') \rangle  \nonumber \\
&\times & {\rm e}^{i(\vec{\ell}'\cdot(\hat{\mathbf{n}}+\vec{\theta})-\vec{\ell}\cdot \hat{\mathbf{n}})}. \nonumber \\
\end{eqnarray}
Now we can use the definition of power spectrum
\begin{eqnarray}
\langle y(\vec{\ell})\kappa(\vec{\ell}') \rangle=(2\pi)^{2}\delta_{\rm 2D}\left( \vec{\ell}-\vec{\ell}'\right)C^{\kappa y}_{\ell}, \label{eq:power-spec}
\end{eqnarray}
and integrate $\vec{\ell}'$ over the 2D Dirac-Delta function, then we have
\begin{eqnarray}
\xi^{\kappa y}(\theta) &=& \int \frac{\der^{2}\hat{\mathbf{n}}}{4\pi} \int 
\frac{\der^{2}\vec{\ell}}{(2\pi)^{2}} C^{\kappa y}_{\ell}{\rm e}^{i\vec{\ell}\cdot \vec{\theta}} \nonumber \\
&=& \frac{1}{(2\pi)^{2}}\int \der \ell \, \ell C^{\kappa y}_{\ell}\int^{2 \pi}_{0}\der \alpha \, {\rm e}^{i \ell \theta \cos\alpha}, \label{eq:xi-kappay-2}
\end{eqnarray}
where in the second line we used the fact that, for statistical isotropy of the Universe, $\langle y(\hat{\mathbf{n}})\kappa(\hat{\mathbf{n}}+\vec{\theta}) \rangle$ does not depend on $\hat{\mathbf{n}}$ thus the integral over $\hat{\mathbf{n}}$ simply gives $4\pi$.

Using the definition of Bessel function
\begin{eqnarray}
J_{n}(x)=\frac{i^{-n}}{2\pi}\int^{2\pi}_{0}\der \theta\, \cos(n\theta){\rm e}^{ix\cos\theta},
\end{eqnarray}
we can simplify Eq.~(\ref{eq:xi-kappay-2}) as
\begin{eqnarray}
\xi^{\kappa y}(\theta) &=& \frac{1}{(2\pi)}\int \der\ell \, \ell C^{\kappa y}_{\ell}J_{0}(\ell \theta) \nonumber \\
&=& \frac{1}{2\pi}\sum_{\ell}\ell C^{\kappa y}_{\ell}J_{0}(\ell \theta),
\end{eqnarray}
where $J_{0}$ is the zero-order Bessel function.

The above calculation is for the two maps without the effect of beam. For {\it Planck} $y-$map and RCSLenS $\kappa$-map, they are smoothed with a Gaussian beam of $\theta_{\rm FWHM}=10$\,arcmin respectively, so the beam function in $\ell$-space is
\begin{equation}
B^{\kappa,y}_{\ell}=\exp\left(-\ell^{2}\sigma^{2}_{\rm b}/2 \right),
\end{equation}
where $\sigma_{\rm b}=\theta_{\rm FWHM}/\sqrt{8\ln 2}=0.00742\left(\theta_{\rm FWHM}/1^{\circ} \right)=1.237 \times 10^{-3}$.
Therefore, the observed correlation function becomes
\begin{eqnarray}
\xi^{\kappa y}(\theta)=\frac{1}{2\pi}\sum_{\ell}\ell C^{\kappa y}_{\ell}J_{0}(\ell \theta)B^{y}_{\ell}B^{\kappa}_{\ell}.
\end{eqnarray}

\subsection{$\xi^{\gamma_{\rm t}y}(\theta)$}

We begin by defining the tangential shear $\gamma_{\rm t}$ on the flat sky~\footnote{The accurate derivation can be done with spin-2 spherical harmonics~\citep{Stebbins96}.}
\begin{eqnarray}
\gamma_{\rm t}(\vec{\theta})=-\gamma_{1}(\vec{\theta})\cos(2\phi)-\gamma_{2}(\vec{\theta})\sin(2\phi), \label{eq:gamma-t}
\end{eqnarray}
where $\vec{\theta}=(\theta\cos\phi,\theta\sin\phi)$ which is the small angle in the flat-sky plane (Fig.~\ref{fig:angle}). $\gamma_{1}$ and $\gamma_{2}$ are the two components of the shear field. On the flat-sky, $\gamma_{1}$ and $\gamma_{2}$ are related to the projected mass density fluctuation in Fourier space, $\kappa(\vec{\ell})$ as
\begin{eqnarray}
\gamma_{1}(\vec{\theta}) &=& \int\frac{\der^{2}\vec{\ell}}{(2\pi)^{2}} \kappa(\vec{\ell})\cos(2 \varphi){\rm e}^{i\vec{\ell}\cdot \vec{\theta}} \nonumber \\
\gamma_{2}(\vec{\theta}) &=& \int\frac{\der^{2}\vec{\ell}}{(2\pi)^{2}} \kappa(\vec{\ell})\sin(2 \varphi){\rm e}^{i\vec{\ell}\cdot \vec{\theta}}, \label{eq:gamma-12}
\end{eqnarray}
where $\varphi$ is the angle between $\vec{\ell}$ and $\hat{\mathbf{e}}_{1}$ (right panel of Fig.~\ref{fig:angle}), i.e. $\vec{\ell}=(\ell \cos \varphi, \ell \sin \varphi)$. Therefore, by combining Eq.~(\ref{eq:gamma-12}) with Eq.~(\ref{eq:gamma-t}), we obtain
\begin{eqnarray}
\gamma_{\rm t}(\vec{\theta})=-\int\frac{\der^{2}\vec{\ell}}{(2\pi)^{2}}\kappa(\vec{\ell})\cos\left[ 2(\phi-\varphi)\right]{\rm e}^{i\ell\theta\cos(\phi-\varphi)}.
\end{eqnarray} 
For purely tangential shears, $\gamma_{\rm t}$ is always positive~\citep{Jeong09}, which allows us to average $\gamma_{\rm t}$ over the ring around the origin to calculate the mean tangential shear $\bar{\gamma}_{\rm t}$
\begin{eqnarray}
\bar{\gamma}_{\rm t}(\vec{\theta}) &=& \int^{2\pi}_{0}\frac{\der \phi}{2\pi}\gamma_{\rm t}(\theta, \phi) \nonumber \\
&=&
- \int^{2\pi}_{0}\frac{\der \phi}{2\pi} \int\frac{\der^{2}\vec{\ell}}{(2\pi)^{2}}\kappa(\vec{\ell}) \nonumber \\
&\times & \cos\left[ 2(\phi-\varphi)\right]{\rm e}^{i\ell\theta\cos(\phi-\varphi)}. \label{eq:gamma-t-theta}
\end{eqnarray}

We now calculate $\xi^{\gamma_{\rm t}y}(\theta)$ as 
\begin{eqnarray}
\xi^{\gamma_{\rm t}y}(\theta) & = & \int\frac{\der^{2}\hat{\mathbf{n}}}{4\pi} \langle
y(\hat{\mathbf{n}})\bar{\gamma}_{\rm t}(\hat{\mathbf{n}}')
\rangle_{\hat{\mathbf{n}}'=\hat{\mathbf{n}}+\vec{\theta}} \nonumber \\
&=& - \int^{2\pi}_{0}\frac{\der \phi}{2\pi} \int\frac{\der^{2}\hat{\mathbf{n}}}{4\pi}\int \frac{\der^{2}\vec{\ell}}{(2\pi)^{2}}\frac{\der^{2}\vec{\ell}'}{(2\pi)^{2}} \langle y(\vec{\ell}')\kappa(\vec{\ell}) \rangle  \nonumber \\
& \times &
\cos\left[2(\phi-\varphi) \right] {\rm e}^{-i\vec{\ell}'\cdot\hat{\mathbf{n}}}{\rm e}^{i\vec{\ell}\cdot(\hat{\mathbf{n}}'=\hat{\mathbf{n}}+\vec{\theta})} \nonumber \\
&=& - \int^{2\pi}_{0}\frac{\der \phi}{2\pi} \int \frac{\der^{2}\hat{\mathbf{n}}}{4\pi} \frac{\der^{2}\vec{\ell}}{(2\pi)^{2}} \cos\left[2(\phi-\varphi) \right] {\rm e}^{i\vec{\ell}\cdot \vec{\theta}} \nonumber \\
&=& - \int^{2\pi}_{0}\frac{\der \phi}{2\pi} \int \frac{\der^{2}\vec{\ell}}{(2\pi)^{2}} \cos\left[2(\phi-\varphi) \right] {\rm e}^{i\ell \theta \cos(\phi-\varphi)}, \nonumber \\
\end{eqnarray}
where in the third equation we have used the definition of power spectrum (Eq.~(\ref{eq:power-spec})), and in the fourth equation we have used the statistical isotropy.

We now carry out the integral as
\begin{eqnarray}
\xi^{\gamma_{\rm t}y}(\theta) & = & - \int^{2\pi}_{0}\frac{\der \phi}{2\pi} \int \frac{\der \ell\, \ell\,\der\varphi}{(2\pi)^{2}} \nonumber \\
& \times & \cos\left[2(\phi-\varphi) \right] {\rm e}^{i\ell \theta \cos(\phi-\varphi)}, \nonumber \\
\end{eqnarray}
and we can re-define the angle $x\equiv \phi-\varphi$ and $y=\phi+\varphi$, we have $\der \phi\der\varphi=\der x\der y$ (Jacobian identity equals to unity), the angles $x$ and $y$ also have the range $[0,2\pi]$, therefore integrate $y$ over, we have
\begin{eqnarray}
\xi^{\gamma_{\rm t}y}(\theta) & = & - \int \frac{\der \ell\,\ell}{(2\pi)^{2}}C^{\kappa y}_{\ell}\left[\int^{2\pi}_{0}\der x \cos(2x){\rm e}^{i\ell\theta\cos x} \right] \nonumber \\
&=& \frac{1}{2\pi} \sum_{\ell}\ell C^{\kappa y}_{\ell}J_{2}(\ell \theta).
\end{eqnarray}
For shear map, it was {\it not} smoothed ($\theta_{\rm FWHM}=0.0$), so the real-space cross-correlation function is
\begin{eqnarray}
\xi^{\gamma_{\rm t}y}(\theta) = \frac{1}{2\pi}\sum_{\ell}\ell C^{\kappa y}_{\ell}J_{2}(\ell \theta)B^{y}_{\ell}.
\end{eqnarray}

\bsp	
\label{lastpage}
\end{document}